\documentclass{aa}  
\usepackage{hyperref}
\usepackage{amsmath}
\usepackage{algorithm}
\usepackage{amsfonts}
\usepackage{mathrsfs}
\usepackage{bm}
\usepackage{rotating}
\usepackage{color}
\usepackage{graphicx}
\usepackage{subfigure}
\usepackage{url}
\usepackage{soul}
\usepackage{natbib}
\bibpunct{(}{)}{;}{a}{}{,}

\def\cha{\textit{Chandra}}
\def\XMM{{XMM-{\it{Newton}}}}
\def\NuSTAR{{\it{NuSTAR}}}
\def\bat{{{\it Swift}-BAT}}
\def\XRT{{{\it Swift}-XRT}}
\def \XSPEC {{\tt{XSPEC}}}

\def \borus {{\tt{borus02}}}

\begin{document}
\title{The properties of the AGN torus as revealed from a set of unbiased NuSTAR observations}
\author{X. Zhao\inst{1,2}\fnmsep\thanks{Email: xiuruiz@clemson.edu} \and S. Marchesi\inst{3,1} \and M. Ajello\inst{1} \and D. Cole\inst{1} \and Z. Hu\inst{1} \and R. Silver\inst{1} \and N. Torres-Alb\`a\inst{1}}
\institute{Department of Physics \& Astronomy, Clemson University, Clemson, SC 29634, USA
\and
Harvard-Smithsonian Center for Astrophysics, 60 Garden Street, Cambridge, MA 02138, USA
\and
INAF-Osservatorio Astronomico di Bologna, Via Piero Gobetti, 93/3, I-40129, Bologna, Italy}

\date{Received NNN; accepted NNN}

 \abstract
{The obscuration observed in active galactic nuclei (AGN) is mainly caused by dust and gas distributed in a torus-like structure surrounding the supermassive black hole (SMBH). However, properties of the obscuring torus of the AGN in X-ray have not been fully investigated yet due to the lack of high-quality data and proper models. In this work, we perform a broadband X-ray spectral analysis of a large, unbiased sample of obscured AGN (with line-of-sight column density 23$\le$log(N$\rm_H$)$\le$24) in the nearby universe which has high-quality archival \NuSTAR\ data. The source spectra are analyzed using the recently developed \borus\ model, which enables us to accurately characterize the physical and geometrical properties of AGN obscuring tori. We also compare our results obtained from the unbiased Compton thin AGN with those of Compton-thick AGN. We find that Compton thin and Compton-thick AGN may possess similar tori, whose average column density is Compton thick (N$\rm_{H,tor,ave}\,\approx$\,1.4 $\times$ 10$^{24}$\,cm$^{-2}$), but they are observed through different (under-dense or over-dense) regions of the tori. We also find that the obscuring torus medium is significantly inhomogeneous, with the torus average column densities significantly different from their line-of-sight column densities (for most of the sources in the sample). The average torus covering factor of sources in our unbiased sample is c$_f$ = 0.67, suggesting that the fraction of unobscured AGN is $\sim$33\%. We develop a new method to measure the intrinsic line-of-sight column density distribution of AGN in the nearby universe, which we find the result is in good agreement with the constraints from recent population synthesis models.}


 \maketitle

%
%
\section{Introduction}
\label{sec:intro} 
Active galactic nuclei (AGN) are one of the most powerful objects in the sky due to the extreme accretion process of the super-massive black holes (SMBHs) in the center of galaxies \citep{Soltan1982,Richstone1998}. The material (gas and dust) surrounding the SMBH not only feeds the central monster but is the origin of the obscuration of the AGN \citep[see][for a recent review]{Hickox18}. Studying the properties of this obscuring material in the AGN is key to understanding the growth of SMBHs.
The AGN unified model \citep{Antonucci1993,Urry_1995} suggests that the obscuring material in the AGN is shaped as an optically thick torus-like structure. The obscuring torus was initially thought to be smooth. However, many recent observations show that the material within the torus is clumpy rather than uniformly distributed \citep[see][for a recent review]{Netzer2015}, e.g., the rapid variable sources and eclipse events found in X-ray observations \citep[e.g.,][]{Risaliti_2002,Markowitz14,Laha_2020} and the observed 10\,$\mu$m silicate features in the Infrared spectra which cannot be explained by a smooth torus \citep[e.g.,][]{Nenkova_2002,Mason09}.

Based on the obscuration of the torus, AGN can be categorized into unobscured when the column density along the line-of-sight is N$_{\rm H,l.o.s}<$ 10$^{22}$\,cm$^{-2}$, and obscured when N$_{\rm H,l.o.s}>$ 10$^{22}$\,cm$^{-2}$. The most obscured sources are called Compton thick (CT-) AGN when N$_{\rm H,l.o.s}>$ 10$^{24}$\,cm$^{-2}$. Moreover, AGN with different levels of obscuration are the main contributors to the cosmic X-ray background \citep[CXB; the diffuse X-ray emission in the universe; e.g.,][]{Maccacaro1991,Madau1994,Comastri1995}. Hard X-ray surveys are more efficient in detecting obscured AGN since high-energy photons can more easily penetrate the dense material surrounding the SMBH. However, only a small number of CT-AGN have been discovered so far \citep[e.g.,][]{risaliti1999,Burlon11,Ricci15,Lanzuisi2018}, due to their large obscuration, which makes it difficult to both detect them and properly measure their column density. Consequently, the uncertainties on the measurement of the intrinsic distribution of the column density of AGN are still significant. The fraction of CT-AGN predicted by AGN population synthesis models \citep[$\sim$30--50\%;][]{gilli07,Ueda14,Buchner2015,Tasnim_Ananna_2019} is much higher than what has been observed so far \citep[$\sim$10--30\%;][]{Burlon11,Ricci15,Lansbury17,Masini18,Zappacosta_2018}. Therefore, the intrinsic column density distribution of AGN is still controversial.

In this work, we perform a broadband X-ray spectral analysis of high-quality (soft and hard) data available for a large, unbiased sample of obscured AGN in the local universe ($z<0.15$). In Section~\ref{sec:selection}, we introduce the criteria that are used to select the unbiased AGN sample and describe the X-ray spectral analysis technique. In Section~\ref{sec:result}, we report the spectral analysis results. In Section~\ref{sec:nh_distribution}, we develop a new method to calculate the intrinsic line-of-sight column density distribution of the AGN in the local universe. We also compare our derived distribution with the results obtained in previous observations and the constraints from different population synthesis models. All reported uncertainties are at 90\% confidence level unless otherwise stated. Standard cosmological parameters are adopted as follows: $<H_0>$ = 70 km s$^{-1}$ Mpc$^{-1}$, $<q_0>$ = 0.0 and $<\Omega_\Lambda>$ = 0.73.

%
\section{Sample selection and Spectral Analysis}
\label{sec:selection}
%
\subsection{Selection Criteria}
The sources presented in this work are selected from the 100-month Palermo $Swift$/BAT catalog\footnote{\url{http://bat.ifc.inaf.it/100m_bat_catalog/100m_bat_catalog_v0.0.htm}} (Marchesi et al., in preparation), which covers 50\% of the sky at the 15--150\,keV flux limit of $\sim$5.4$\times$10$^{-12}$\,erg\,cm$^{-2}$\,s$^{-1}$. The selection criteria are as follows:
\begin{enumerate}
\item \ul{Sources with line-of-sight column density\footnote{For sources also detected in the BAT 70-month catalog, we use the spectral properties measured in \citep{Ricci2017}. We measured their spectral properties using the Swift-BAT data and soft X-ray data for the rest of the sources.} (N$\rm _{H,l.o.s}$) between 10$^{23}$ and 10$^{24}$\,cm$^{-2}$.} To characterize the physical and geometrical properties of the obscuring material around the SMBH, one needs a clear signal of the reprocessed component of the obscuring torus overcoming the line-of-sight component. CT-AGN, which are ideal for studying the obscuring torus, are significantly biased against when sampling. Thus here we select heavily obscured Compton thin AGN, which are least biased when studying the obscuring torus. In sources with N$\rm _{H,l.o.s}$ $<$10$^{23}$\,cm$^{-2}$, the contribution of the reprocessed component to the overall AGN emission is negligible \citep[$<$5\% at 2--10\,keV;][]{Borus}, rendering the derivation of the tori properties a difficult process in those sources. Thus, we select Compton thin sources with N$\rm _{H,l.o.s}$ $>$10$^{23}$\,cm$^{-2}$ in our study of the AGN torus. We analyzed the nearby CT-AGN selected in \bat\ catalog in a separate set of papers \citep[][Marchesi et al. in prep.]{Marchesi2018,Marchesi_2019,Zhao_2019_2,Zhao_2019_1}.

\item \ul{Available NuSTAR data.} \NuSTAR\ data are instrumental to properly characterize the properties of heavily obscured AGN in the local universe due to the significant suppression of their spectra at soft X-rays \citep[see, e.g.,][]{Civano_2015,Marchesi2018}.
\end{enumerate}

The obscuring material that reprocesses the X-ray emissions from the `hot corona’ is mainly related to the dusty torus ($<$10\,pc) proposed by the AGN unified model and to the interstellar medium (ISM; $>$kpc) of the host galaxy \citep[see,][for recent reviews]{Netzer15,Naure_astro2017,Hickox18}. The typical column density N$\rm _H$ toward the nucleus decreases with size scale as R$^{-2}$ (where R is the distance from the central SMBH). Thus the N$\rm _H$ from the galaxy-wide scale is thought to be N$\rm _H$ $<$10$^{23}$\,cm$^{-2}$, except for galaxy mergers \citep[e.g.,][]{Di-Matteo:2005aa} or high-redshift quasars where the galaxies are rich in gas \citep[][]{Circosta_2019}. The dusty gas in the compact torus can produce obscuration up to N$\rm _H$ $\approx$10$^{25}$\,cm$^{-2}$. Our selected sources are heavily obscured (N$\rm _H$ $>$10$^{23}$\,cm$^{-2}$) and are in the local universe ($z<$0.15), and no evidence has been found for merger events. Therefore, we assume that the obscuring material that reprocesses the X-ray emission in our sources is mainly from the AGN torus.

A total of 93 out of $\sim$1000 AGN in the BAT catalog have been selected and analyzed in this work. The information about the observations used when analyzing each source is listed in Appendix. The median redshift of the sources in our finalized sample is $\left<z\right>$ = 0.02776 (i.e., the median distance is $\left<d\right>$ $\sim$122\,Mpc).

%
\subsection{Spectral Analysis}
\label{section:model}
%
We perform broadband (1--78\,keV) X-ray spectral analysis of all 93 sources in our sample. \NuSTAR\ (3--78\,keV) data provide coverage in the hard X-rays. For the soft X-ray band, we use archival \XMM\ data when they are available (1--10\,keV; 48 sources). We use archival \cha\ data (1--7\,keV; 19 sources) when \XMM\ data are not available. We use \XRT\ data (1--10\,keV; 26 sources) when neither \XMM\ nor \cha\ data are available. The details of the data reduction are listed and described in Appendix \ref{sec:data_reduction}. 

The spectra are fitted using \texttt{XSPEC} \citep{Arnaud1996} version 12.10.0c. The photoelectric cross-section is from \cite{Verner1996}; the element abundance is from \citet{Anders1989} and metal abundance is fixed to Solar; the Galactic absorption column density is obtained using the \texttt{nh} task \citep{Kalberla05} in \texttt{HEAsoft} for each source. The $\chi^2$ statistic is adopted when \XMM\ data and \cha\ data are used, while C statistic \citep{Cash1979} is used when \XRT\ data are applied due to the low quality of the \XRT\ spectra and the limited number of counts in each bin.

The spectra of heavily obscured AGN are complicated by the emergence of the reprocessed component, including the Compton scattering and fluorescent emission lines, which are buried by the line-of-sight component in the unobscured AGN spectra. However, despite adding the difficulty of characterizing the spectra, this reprocessed component becomes a perfect tool to estimate the physical and geometrical properties of the obscuring material surrounding the central SMBH. In this work, following \citet{Zhao_2020}, we analyze the spectra of the sources in our sample using the self-consistent Borus model \citep[][]{Borus}, which optimizes the exploration of the parameter space and has been intensively used to characterize heavily obscured AGN.
 
 \begin{figure*} 
\begin{minipage}[b]{.5\textwidth}
\centering
\includegraphics[width=1\textwidth]{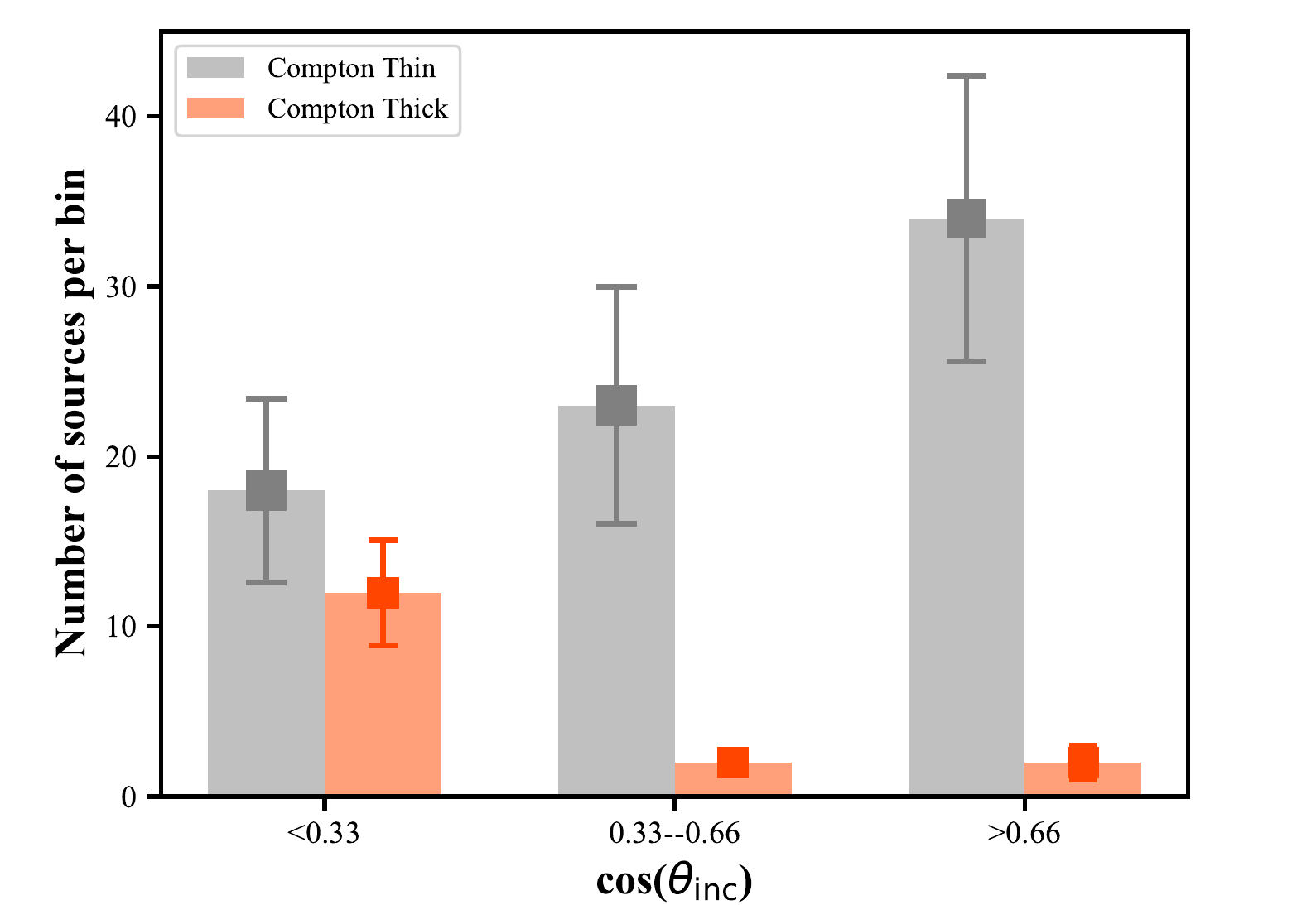}
\includegraphics[width=1\textwidth]{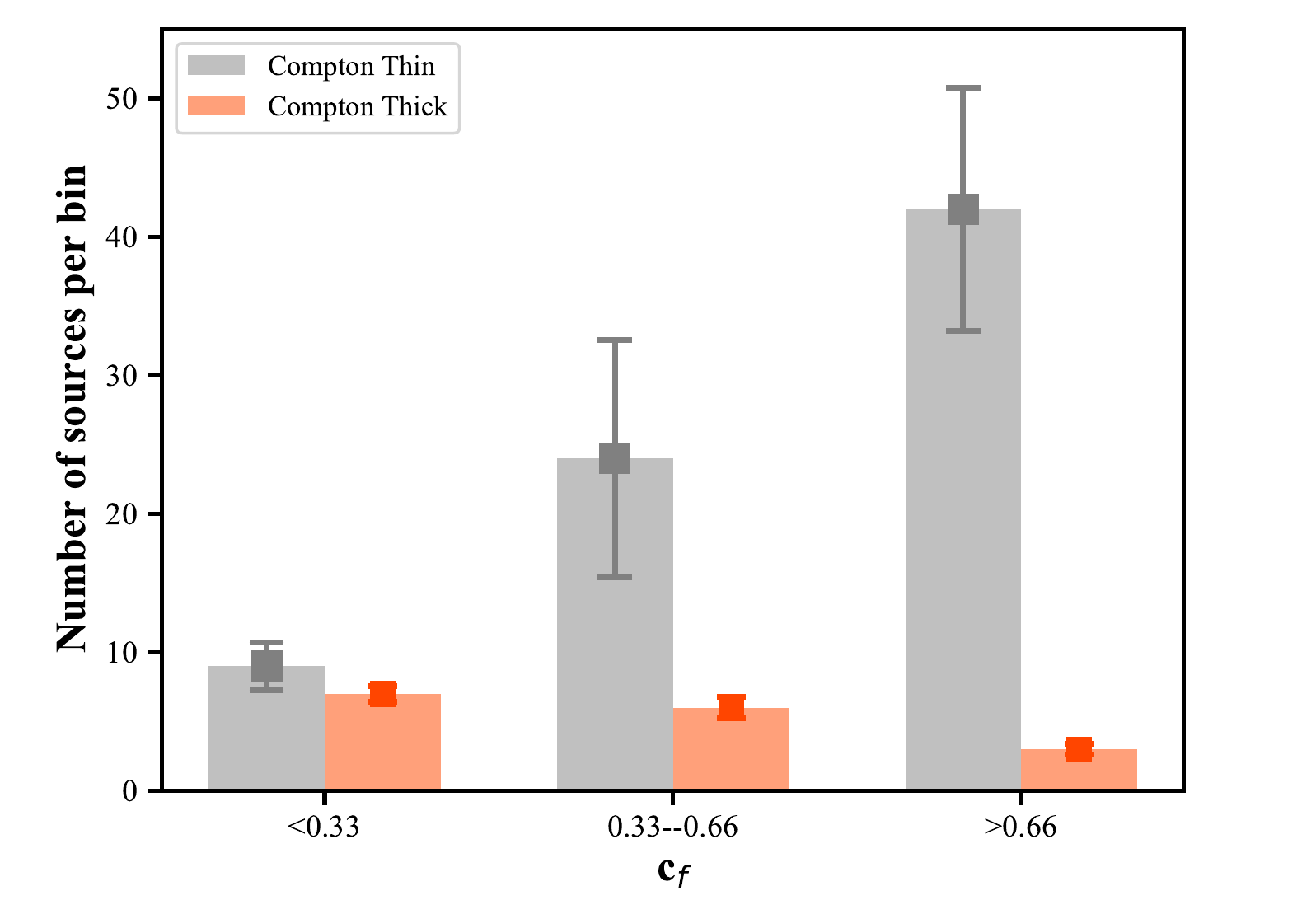}
\end{minipage}
\begin{minipage}[b]{.5\textwidth}
\centering
\includegraphics[width=1\textwidth]{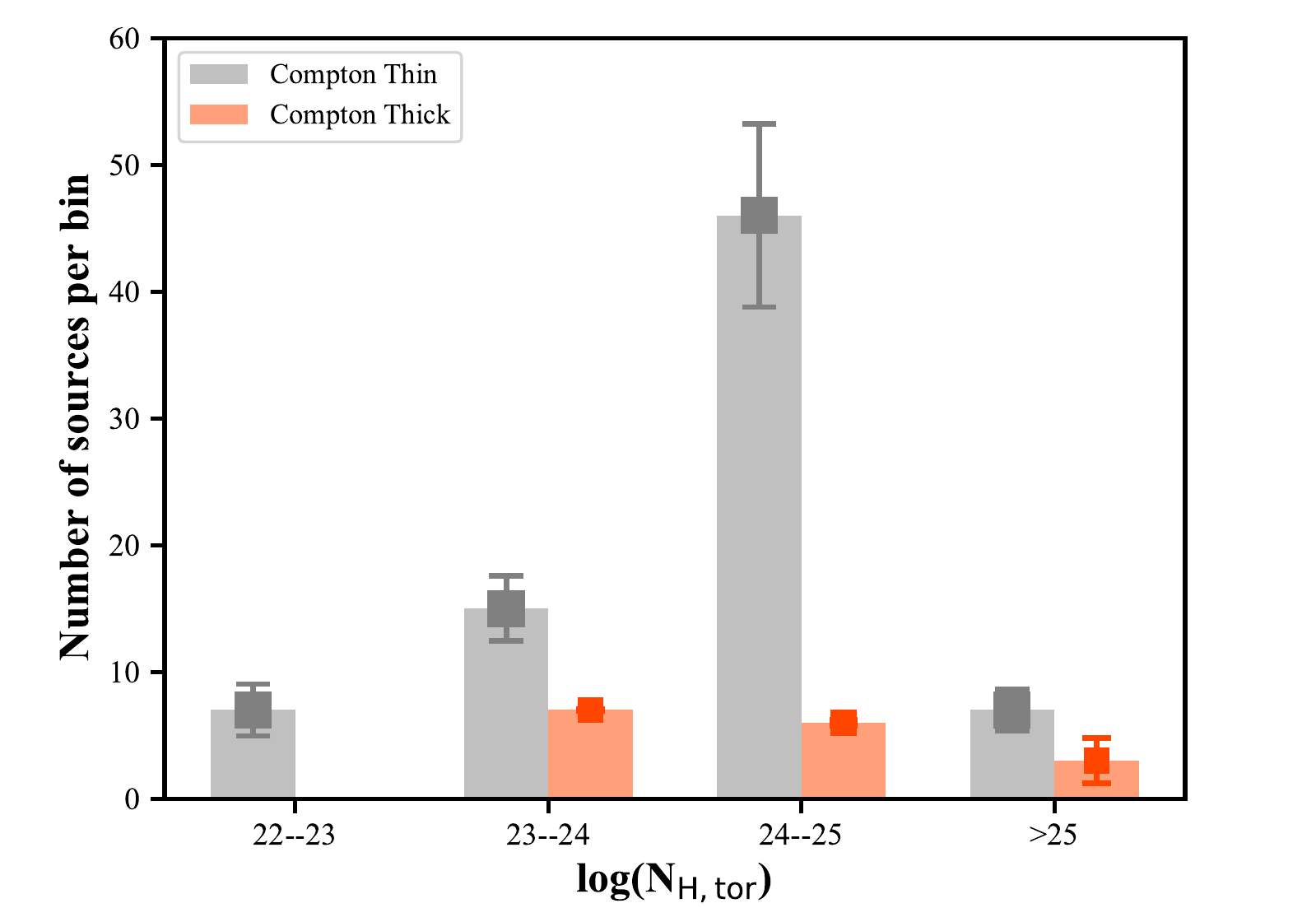}
\includegraphics[width=1\textwidth]{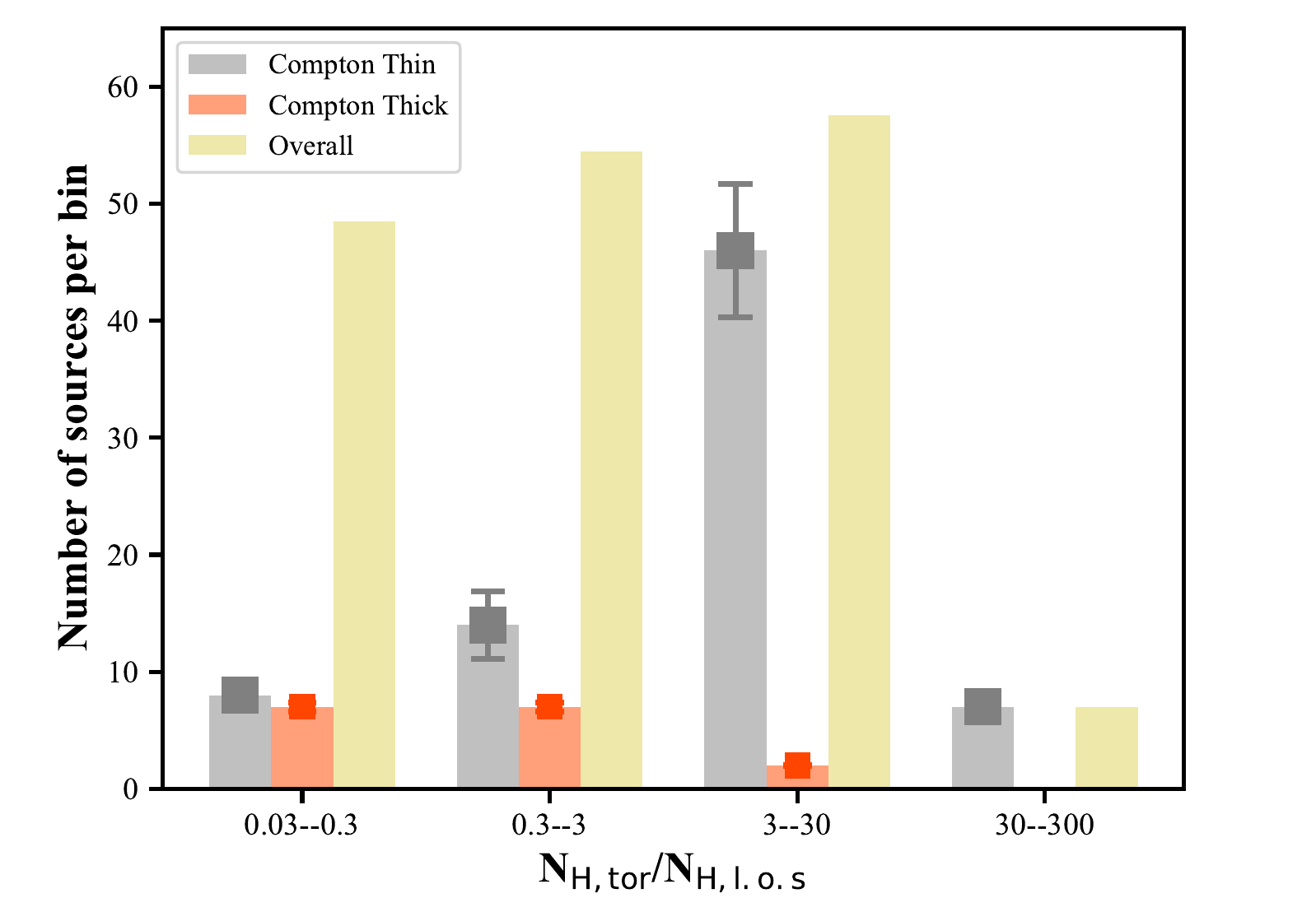}
\end{minipage}
\caption{The figures from left to right in the first row and second row are the number of sources with a specific range of the inclination angles, cos($\theta_{\rm inc}$), torus column densities, log(N$_{\rm H,tor}$), torus covering factors, $c \rm_f$, and the torus column density contrast ratio, N$_{\rm H,tor}$/N$_{\rm H,l.o.s}$, of Compton thin sources in our sample (grey histogram) and CT sources from M19 (orange histogram), respectively. Cos($\theta_{\rm inc}$) $\sim$0.05 is when the AGN is observed in `edge-on' direction and cos($\theta_{\rm inc}$) $\sim$0.95 is when the AGN is observed `face-on'. The error bar is at 90\% confidence level. The overall distribution of torus column density contrast ratio after rescaling is plotted in yellow. We do not include the 15 sources fitted with only a line-of-sight component and a scattered component here.}
\label{fig:bar_figure}
\end{figure*}   

The complete Borus model used in the spectral analysis consists of three components: 

{\bf A)} a reprocessed component, produced by the obscuring material near the SMBH, including the Compton scattered continuum and fluorescent lines, characterized by \borus. \borus\ assumes a sphere with conical cutouts at both poles \citep{Borus}, approximating a torus with an opening angle which can vary in the range of $\theta_{\rm Tor}$ = [0--84]$^\circ$, corresponding to a torus covering factor, $c_f$ = cos($\theta_{\rm Tor}$) = [1--0.1]. The inclination angle, which is the angle between the axis of the AGN and the observer line-of-sight, is also a free parameter ranging from $\theta_{\rm inc}$ = [18--87]$^\circ$, where $\theta\rm_{obs}$ = 18$^\circ$ is when the AGN is observed ``face-on'' and $\theta\rm_{obs}$ = 87$^\circ$ is observed ``edge-on". The average column density of the obscuring torus, can vary in the range of log(N$_{\rm H,tor}$) = [22.0--25.5].

{\bf B)} a line-of-sight component or the absorbed intrinsic continuum, described by a cut-off power law, denoted by \texttt{cutoffpl} in \XSPEC, multiplied by an obscuring component, including both the photoelectric absorption (\texttt{zphabs}) and the Compton scattering (\texttt{cabs}) effects. It is worth noting that the torus column density in the reprocessed component is decoupled from the line-of-sight column density: the torus column density is an average property of the obscuring torus. In contrast, the line-of-sight column density represents a quantity that is along our line-of-sight and could vary when observed at different epochs. 

The line-of-sight column density N$_{\rm H,l.o.s}$ can be significantly different from N$_{\rm H,tor}$, which can be used as an approximation of the non-uniform (clumpy) distribution of matter surrounding the SMBH \citep[see discussion in Sec.~4.2 of][]{Borus}. \citet{Buchner19,Tanimoto_2019}, which assumes a clumpy torus scenario, provides a more realistic description of the obscuring material surrounding the accreting SMBH. However, they do not provide information about the torus' global average properties, e.g., the torus average column density, which is a significant focus of this paper. Moreover, the \citet{Tanimoto_2019} model has not been publicly available yet.
 
{\bf C)} a scattered component, modeling the fractional AGN emission that is scattered rather than absorbed by the obscuring material. The scattered component is characterized by an unabsorbed cut-off power law multiplied by a constant. The fractional unabsorbed continuum is usually less than 5--10\% of the intrinsic continuum \citep[see, e.g.,][]{Noguchi_2010,marchesi2016} and accounts for the AGN emission at low-energy below a few keV.

In the process of modeling the spectra, we tie the photon indices, $\Gamma$, the cut-off energy, $E_{\rm cut}$ and the normalization, $\rm norm$, of the intrinsic continuum, the reprocessed component, and the fractional unabsorbed continuum together, assuming that the three components have the same origin. The photon index in \borus\ ranges in [1.4--2.6], thus the photon index in the cut-off power-law varies between 1.4 and 2.6 as well.

The Borus model is used in the following \XSPEC\ configuration:
\begin{equation}\label{eq:Borus}
\begin{aligned}
Model =\,constant_1*phabs*(borus02+&\\
zphabs*cabs*cutoffpl+constant_2*cutoffpl)&
\end{aligned}
\end{equation}
where \texttt{constant$_1$} is the cross-calibration between \NuSTAR\ and the soft X-ray observatories, i.e., \XMM, \cha\ and \XRT; \texttt{phabs} models the Galactic absorption from our Galaxy; \texttt{constant$_2$} is the fraction of the unabsorbed continuum in the scattered component.

The cut-off energy is fixed at $E_{\rm cut}$ = 500\,keV for all sources, except for ESO 383-18, whose cut-off energy is found to be $E_{\rm cut}$ $<$20\,keV. The spectra of nine sources (ESO 263-13, Mrk 3, NGC 835, NGC 2655, NGC 4102, NGC 4258, NGC 4507, NGC 5728 and NGC 7319) all show strong non-AGN thermal emission around 1\,keV, which may be caused by the star formation process and/or diffuse gas emission. We use \texttt{mekal} \citep{mekal} to model this non-AGN thermal contribution: the temperature and the relative metal abundance in \texttt{mekal} are both left free to vary. To better constrain the parameters of the \texttt{mekal} model, we fit the spectra of these sources down to 0.3\,keV. When analyzing the spectra of 3C 445, we add a few gaussian lines to model different emission lines. The relative iron abundance of the reprocessed component, A$\rm _{Fe}$, is fixed to 1 (the solar value), except for three sources (3C 445, 3C 452 and ESO 383-18), where a significant improvement in spectral fit ($>3\,\sigma$ using \texttt{ftest} in \texttt{XSPEC}) has been found when A$\rm _{Fe}$ is free to vary. The best-fit iron abundances are A$_{\rm Fe,3C\,445}$ = 0.32$_{-0.08}^{+0.05}$, A$_{\rm Fe,3C\,452}$ = 0.41$_{-0.05}^{+0.05}$, and A$_{\rm Fe,ESO\,383-18}$ = 0.19$_{-0.05}^{+0.07}$.

We note that the obscuring material in the AGN might also originate from the polar dust outflow as revealed by Infrared observations \citep[e.g.,][]{Tristram07}. However, the X-ray observations are less affected by the polar outflow since the gas density in the polar outflow is much less than that in the torus \citep{Wada_2016}. Indeed, the covering factors measured in X-ray are systematically smaller than those measured in infrared \citep[see, Fig.2 in][]{Tanimoto_2019}. This suggests that the typical size of the X-ray obscuring material is much smaller (i.e. the torus) than the typical sizes of the IR emission (i.e. torus + polar dust). Therefore, in the following discussion, we assume that absorption measured in the X-ray spectra of the sources in our sample is mainly due to the obscuring material in the torus, except for a few cases discussed in Section~\ref{sec:result}.

%
\section{Results} 
\label{sec:result}
%
\begin{figure*} 
\includegraphics[width=.5\textwidth]{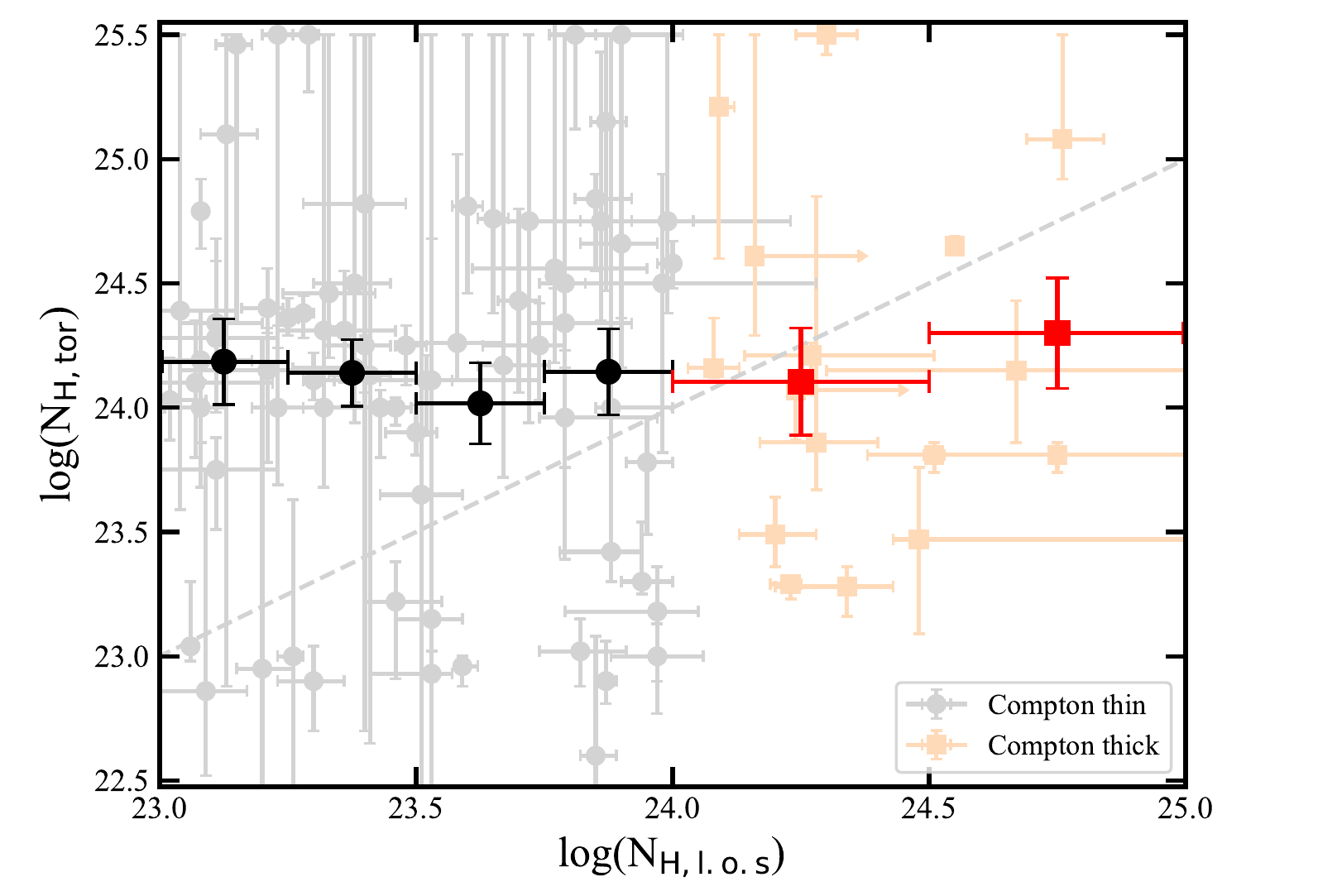}
\includegraphics[width=.5\textwidth]{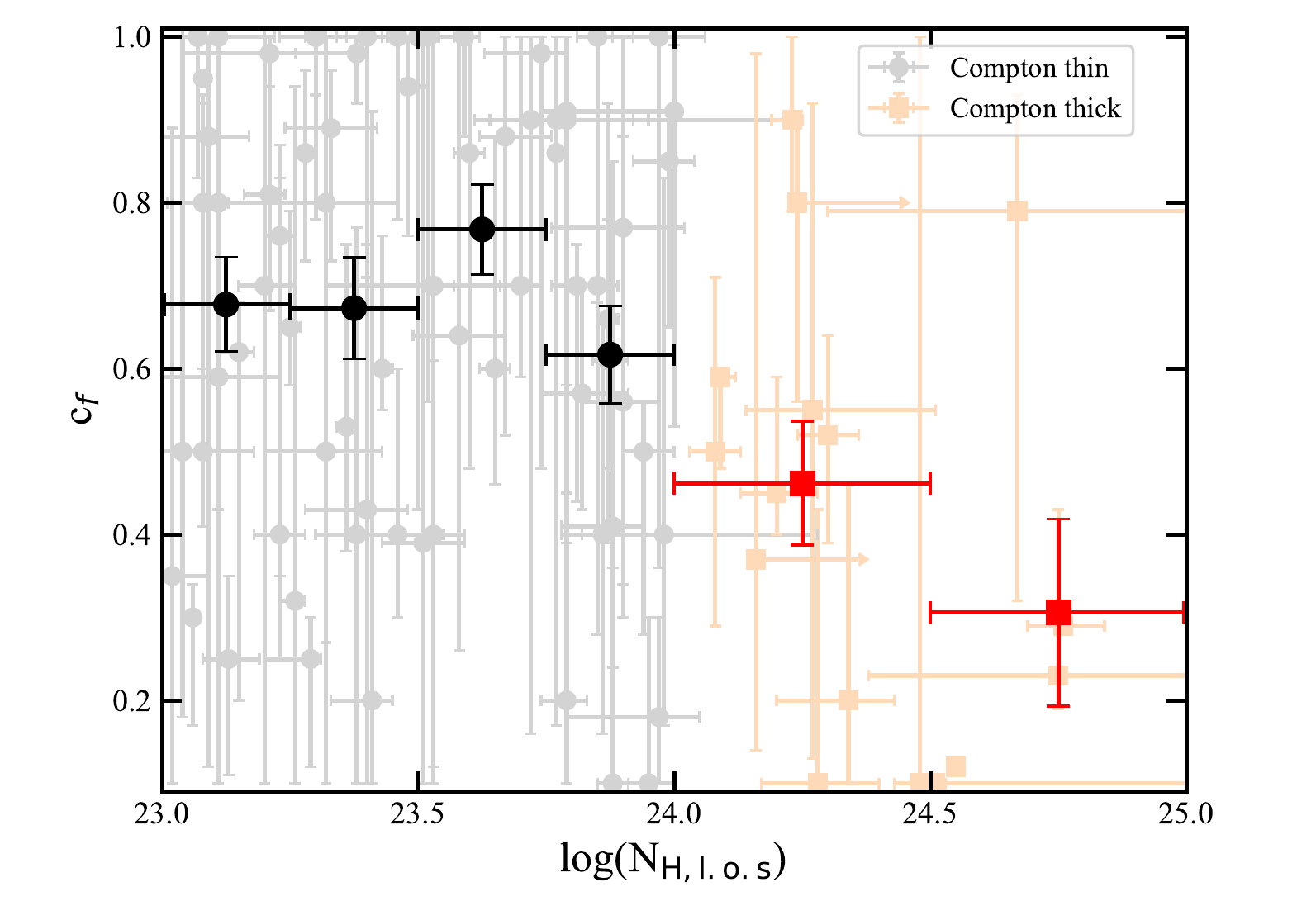}
\caption{\normalsize Torus column density log(N$_{\rm H,tor}$) (left) and torus covering factor c$_f$ (right) as a function of line-of-sight column density, log(N$_{\rm H,l.o.s}$). Compton thin sources are plotted in light grey points, while CT sources are plotted in light red squares. The average and 1\,$\sigma$ standard error of the torus column densities and torus covering factors in different line-of-sight column density bins are plotted in black points for Compton thin sources and red squares for CT sources, respectively. In the left panel, N$_{\rm H,tor}$ = N$_{\rm H,l.o.s}$ is plotted as a light grey dashed line. The low covering factors of the CT sources are due to the bias explained in Sec.~\ref{sec:result}.}
\label{fig:NHZ-NHS}
\end{figure*}   

The spectra of 93 sources and their best-fit models can be found online\footnote{\url{https://science.clemson.edu/ctagn/project/}}. The best-fit results of the spectral analysis, e.g., line-of-sight column density, N$_{\rm H,l.o.s}$, torus average column density, N$_{\rm H,tor}$, cosine of the inclination angle, cos($\theta_{\rm inc}$), torus covering factors, $c_{\rm f}$, 2--10\,keV flux, Flux$_{2-10}$, and 2--10\,keV intrinsic (absorption-corrected) luminosity, L$\rm _{int,2-10}$ are reported in Table~\ref{Table:resultsXMM}, \ref{Table:resultsCha}, \ref{Table:resultsXRT} when \XMM, \cha, and \XRT\ data are used, respectively. For 15 sources in our sample, the contributions of their reprocessed emission to the overall spectra are marginal, such that the parameters of the reprocessed component, e.g., the torus average column density, inclination angle, and torus covering factor, cannot be constrained at all. Therefore we analyze their spectra using only the line-of-sight component and the scattered component for those sources. Eight sources in our sample are found to have line-of-sight column density N$_{\rm H,l.o.s}$ $<$10$^{23}$\,cm$^{-2}$ or N$_{\rm H,l.o.s}$ $>$10$^{24}$\,cm$^{-2}$, thus these sources are excluded from our sample. The spectra and best-fit model predictions of four representative sources have been plotted in Fig.~\ref{fig:spectra}.

Flux variations are commonly found in the X-ray spectra of AGN, especially when multiple observations are taken over time-scales that vary from weeks to years \citep[e.g.,][]{Guainazzi02,Risaliti02,Markowitz14}. Such variabilities are commonly explained by either the fluctuation of the AGN intrinsic emission \citep[e.g.,][]{NANDRA2001295}, or the variation in the absorbing column density along the ``line-of-sight'' \citep[see, e.g.,][]{Risaliti_2002,Bianchi12}. Within the 93 sources in our sample, 20 sources are semi-simultaneously observed by \NuSTAR\ and the soft X-ray observatories. In the remaining 73 sources, 43 ($\sim$59\%) are measured with significant flux variations between the soft and hard X-ray observations (2--10\,keV flux variation $>$20 percent). Within these 43 sources, 16 mainly show intrinsic emission variation; 15 sources experienced N$_{\rm H,l.o.s}$ variation; and the flux variabilities of 12 sources can be explained only by considering both intrinsic emission and N$_{\rm H,l.o.s}$ variations. The details of the variability analysis are reported in Appendix~\ref{sec:variability}. The best-fit results of the sources with flux variability are summarized in Table~\ref{Table:Variability}. It is worth noting that if a source had multiple soft X-ray observations, we chose the one taken closest to their \NuSTAR\ observation time. Therefore, the fraction of the sources in our sample that experienced flux variabilities might be even higher if all soft X-ray observations are taken into account. Although, we cannot exclude the possibility that the N$_{\rm H,l.o.s}$ measured with different instruments may have different values (apparent variability) due to different bandpasses, energy resolution, and effective areas. The N$_{\rm H,l.o.s}$ reported in Table~\ref{Table:resultsXMM}, \ref{Table:resultsCha}, \ref{Table:resultsXRT} are those measured in the \NuSTAR\ observing epoch, and we will use them in the following discussions.

\citet[][hereafter M19]{Marchesi_2019} reported the CT-AGN in the nearby universe, which are also selected from the BAT 100-month catalog and have \NuSTAR\ data. The sources were analyzed using \borus\ model as well, with the distinction that the inclination angle is fixed at `edge-on' direction when fitting the spectra. We re-analyze these CT sources using the same model presented in Section~\ref{section:model} and let the inclination angle free to vary when fitting the spectra. Thus we could compare the properties of the obscuring torus of CT sources with those of the Compton thin sources in our sample. The best-fit re-analysis results of the CT sources in the M19 sample are reported in Table~\ref{Table:resultsCT}. Four CT sources in the M19 sample are found to be Compton thin in our re-analysis. Therefore, we add those four sources into our Compton thin sample, which in the end includes 89 sources in total\footnote{The median photon indices of the 89 sources is $\left<\Gamma \right>$ = 1.67.}. 

Although CT-AGN are an ideal sample to study the obscuring torus due to their significantly suppressed line-of-sight emission, thus showing clearly the reprocessed emission from the torus, we need to keep in mind that the strong bias against the detection of CT-AGN, which makes the observed CT sample incomplete. Indeed, the detected fraction of CT-AGN ($f_{\rm CT}$) in the BAT sample decreases significantly as the distance (redshift) increases \citep[$f_{\rm CT}\approx$ 31\% at 10\,Mpc and $f_{\rm CT}\approx$ 4\% at 100\,Mpc;][]{Ricci15}, suggesting that the detection of CT-AGN is significantly biased, especially for low-luminosity CT-AGN at large distance. Therefore, the torus properties derived from the CT sample in this work may not represent the torus properties of the whole CT-AGN population.

The main results of the spectral analysis are:
\begin{itemize}
\item The Compton thin sources in our sample are evenly observed in every direction considering the uncertainties as shown in Fig.~\ref{fig:bar_figure} (upper left), suggesting that our BAT selected Compton thin AGN sample is an \emph{unbiased} sample. In contrast, most of the CT sources in the M19 sample are observed `edge-on', which is in agreement with the material distribution formalism used in the clumpy torus model that the obscuring clumps are populated in the `edge-on' direction \citep[e.g.,][]{Nenkova08,Buchner19,Tanimoto_2019}.

\item The torus column densities of the majority of Compton thin sources and CT-AGN are in the $>$10$^{24}$\,cm$^{-2}$ range as presented in Fig.~\ref{fig:bar_figure} (upper right), suggesting that CT reflectors are commonly found in both Compton thin and CT-AGN. A similar result was also reported in, e.g., \citet{Buchner19}.

\item The average torus column density is similar for both Compton thin and CT-AGN, independent on the observing angle, i.e., log(N$\rm_{H,tor,ave}$) $\sim$24.15, as shown in Fig.~\ref{fig:NHZ-NHS} (left). 

\item We notice that the torus average column densities of Compton thin sources are generally larger than their line-of-sight column density, suggesting that Compton thin AGN are usually observed through an {\bf under-dense} region in their tori. The torus average column density of CT-AGN is always smaller than their line-of-sight column density, suggesting that CT-AGN are observed through an {\bf over-dense} region in their tori. Although we cannot exclude the possibility that the additional obscuration along the line-of-sight of sources (especially when observed face-on) with N$_{\rm H,tor}$ $<$ N$_{\rm H,l.o.s}$ is due to polar outflows \citep{H_nig_2013}.

\item Compton-thin and CT-AGN have (statistically) different covering factors, with the former having larger ($c\rm_f>$\,0.5) covering factors while the latter having smaller ($c\rm_f<$\,0.5) covering factors, as shown in Fig.~\ref{fig:bar_figure} (bottom left). Interestingly, the average torus covering factor of Compton thin AGN is c$_{f\rm,ave}$ $\sim$0.67. In contrast, the average torus covering factor decreases significantly when the line-of-sight column density reaches the CT regime, as shown in Fig.~\ref{fig:NHZ-NHS} (right). The low-$c\rm_f$ found in the CT sample might reflect the fact that the CT sample is significantly biased. The detected CT-AGN in the BAT sample are the intrinsically most luminous sources, which typically have a lower $c\rm_f$ compared with the intrinsically fainter AGN \citep{Burlon11,Ricci:2017aa,Marchesi_2019}, suggesting that BAT samples only the low-$c\rm_f$ CT-AGN even in the nearby Universe. Indeed, \citet{Ricci15} measured a c$_f\approx$ 0.70 for AGN in the local Universe using a statistical method (the fraction of the absorbed AGN is used as a proxy of the mean torus covering factor of the AGN in the sample). Their result is in good agreement with the average covering factor measured in our unbiased Compton thin sample.

\item The obscuring material in the torus of AGN is \emph{significantly inhomogeneous}. The torus average column densities of the majority of Compton thin and CT-AGN are at least three times different (larger or smaller) from their line-of-sight column densities as shown in Fig.~\ref{fig:bar_figure} (bottom right). Moreover, for $\sim$30\% of sources in the sample, their torus average column densities are more than ten times different from their line-of-sight column densities. A similarly inhomogeneous torus result is also found in \citet{Laha_2020}, which studied a sample of 20 Compton thin AGN.

\item The distributions of the torus column density contrast ratio (CR = N$_{\rm H,tor}$/N$_{\rm H,l.o.s}$ which shows how inhomogeneous the torus is) observed in Compton thin and CT-AGN are different (see, Fig.~\ref{fig:bar_figure} bottom right). We consider that the overall distribution of the torus column density contrast ratio of different AGN types should be a combination of the CR distribution of both Compton thin and CT-AGN. To combine the two distributions, we rescale the torus column density contrast ratio distribution of CT-AGN before we produce the overall distribution since we are biased against the detection of CT-AGN, i.e., we need to consider the intrinsic number of Compton thin and CT-AGN. Here, we assume that the number of Compton thin AGN and CT-AGN are similar, which is in agreement with previous works and what we find in this work, as we will further discuss in Section~\ref{sec:nh_distribution}. After rescaling, we find that the torus column density contrast ratio is rather flat over CR = [0.03,30] (Fig~\ref{fig:bar_figure} bottom right), suggesting that for a given torus column density, an AGN can be observed with different line-of-sight column densities with equal probabilities.

\end{itemize}

In the Compton thin sample, the effect of the biases discussed above are much weaker than in the CT sample, since the reprocessed component does not dominate the hard band emission of Compton thin AGN. Therefore, the tori properties measured in the Compton thin sample are least biased against. We will only use these measurements to derive the intrinsic $\rm N_{H, l.o.s.}$ distribution of AGN in the local universe in Section~\ref{sec:nh_distribution}.

\begin{figure} 
\centering
\includegraphics[width=0.5\textwidth]{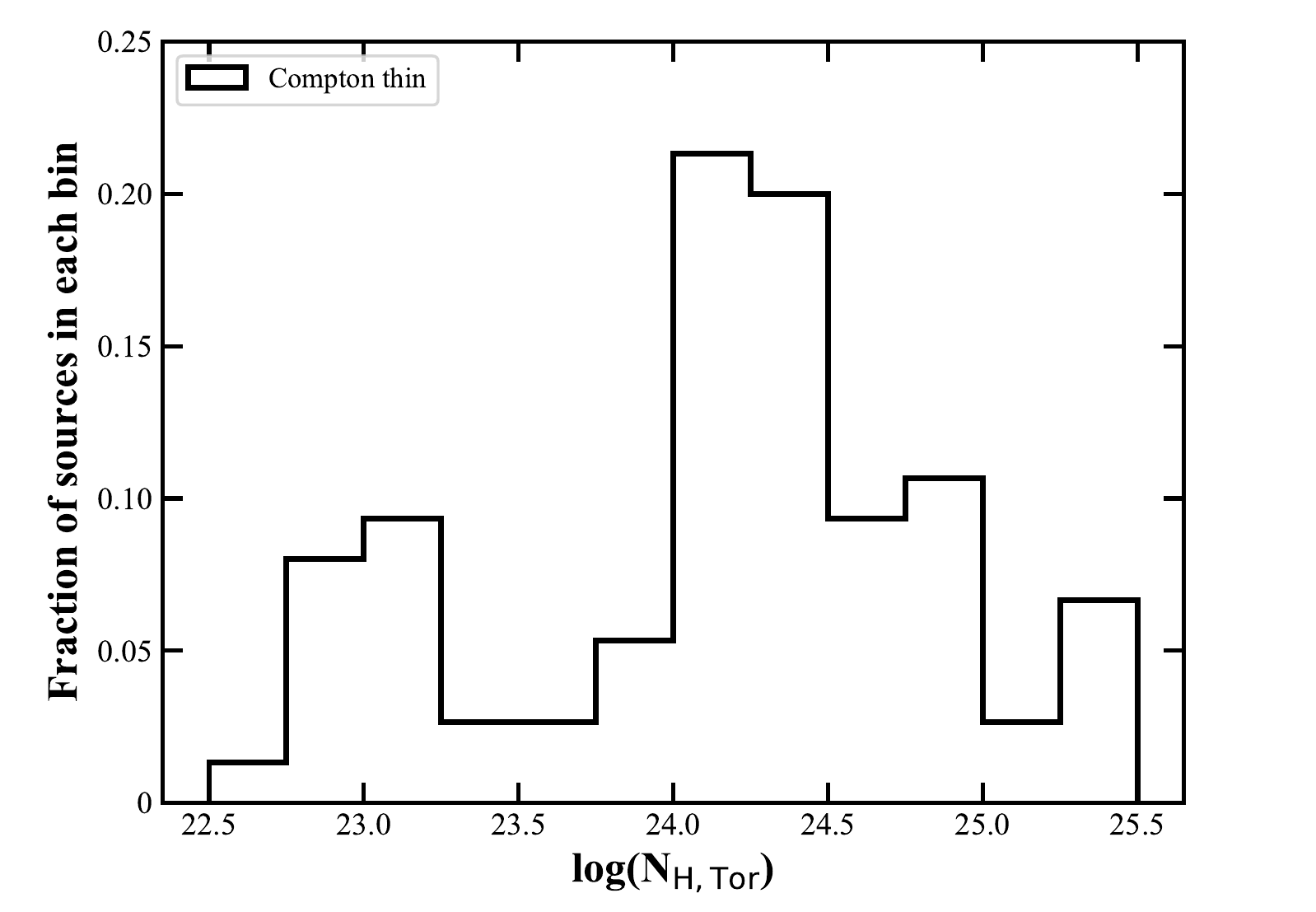}
\caption{\normalsize Torus average column density distribution of the sources in our unbiased Compton thin AGN sample. The histogram describes the same sample as in the upper right panel of Fig.~\ref{fig:bar_figure} but with different grouping bins.}
\label{fig:nhs}
\end{figure}   

%
\section{Intrinsic line-of-sight Column Density Distribution of AGN} 
\label{sec:nh_distribution}
%
\begin{figure*} 
\begin{minipage}[b]{.5\textwidth}
\centering
\includegraphics[width=1\textwidth]{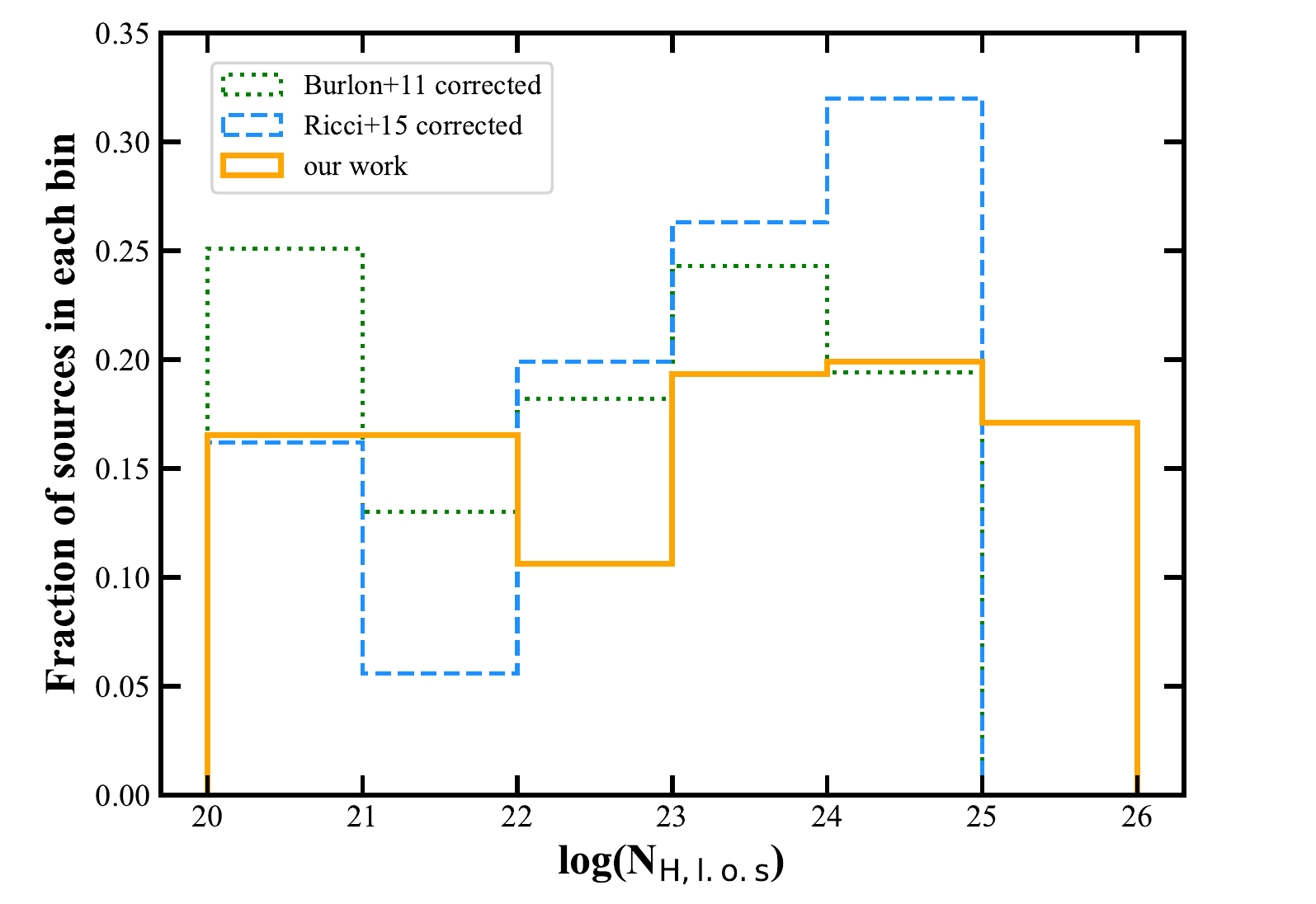}
\end{minipage}
\begin{minipage}[b]{.5\textwidth}
\centering
\includegraphics[width=1\textwidth]{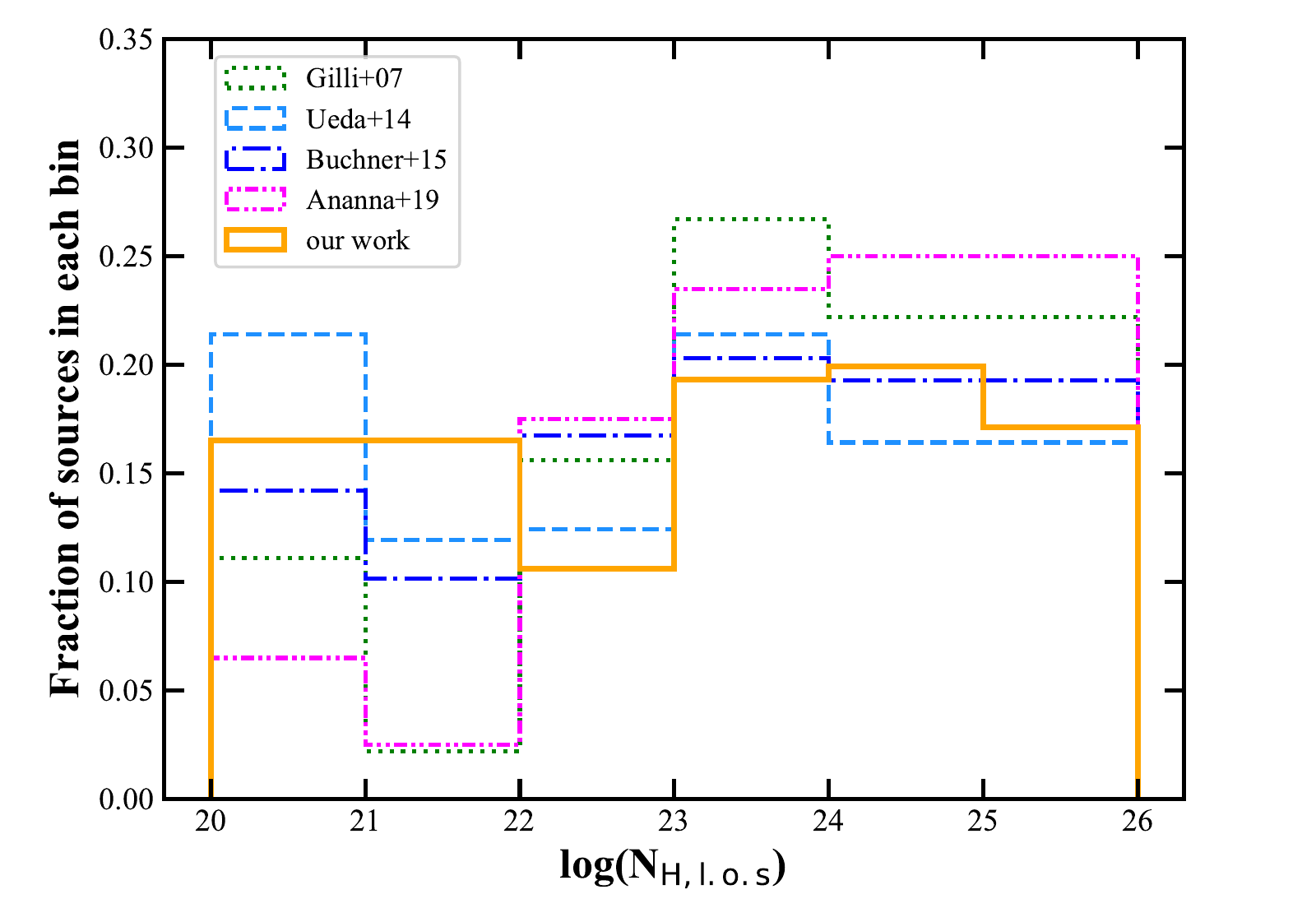}
\end{minipage}
\caption{Comparison between the intrinsic line-of-sight column density distribution derived from our new developed method with previous observational results (left) and those used in different population synthesis models (right). The extremely obscured sources with $\rm log(N_{H,l.o.s})$ $>$26 are included in the 25 $<\rm log(N_{H,l.o.s})$ $<$26 bin.}
\label{fig:nh_compare}
\end{figure*}   

The line-of-sight column density distribution of AGN in the local universe has been measured in recent years \citep[e.g.,][]{Burlon11,Ricci15}. However, the observed line-of-sight column density distribution is significantly selection-biased. Thus the intrinsic distribution of the AGN line-of-sight column density depends heavily on the absorption correction used particularly for $\rm N_{H, l.o.s.}>10^{23.5}$\,cm$^{-2}$. To investigate the actual intrinsic line-of-sight column density distribution of AGN in the local universe, we develop a  Monte Carlo method based on the measured properties of the obscuring tori in our unbiased sample from Section~\ref{sec:result}. Here we assume that different types of AGN possess tori with similar properties \citep{Antonucci1993,Urry_1995}. The details of our method are as follows:
\begin{enumerate}
\item We draw randomly a torus average column density $\rm N_{H,tor}$ from the distribution shown in Fig.~\ref{fig:nhs};
\item We then assign a random CR in the range of [0.03--30] to each $\rm N_{H,tor}$ (only a small fraction of sources have extremely inhomogeneous torus with CR outside of this range), thus the line-of-sight column density of each torus is $\rm N_{H,l.o.s}$ = $\rm N_{H,tor}$/CR;
\item we assume a 33\% fraction of unobscured AGN since the average covering factor of the torus in our sample is c$_{f\rm,ave}$ $\sim$0.67 thus the line-of-sight of 33\% AGN do not intersect their torus. We presume that the distribution of $\rm N_{H,l.o.s}$ between 10$^{20}$\,cm$^{-2}$ and 10$^{22}$\,cm$^{-2}$ is flat.
\end{enumerate}

A  discrepancy of the intrinsic line-of-sight column density distribution between the previous observational results obtained using  \bat\ \citep[e.g.,][]{Burlon11,Ricci15} and constraints from different population synthesis models \citep[e.g.,][]{gilli07,Ueda14,Buchner2015,Tasnim_Ananna_2019} still exists. We compare our derived intrinsic line-of-sight column density distribution with their results \footnote{The reported observational column density distributions are those which have been corrected for selection bias and for the bias against detecting extremely absorbed sources. The average redshift of the sources in the samples used in \citet{Burlon11} and \citet{Ricci15} are $\left<z\right>$ = 0.03 and 0.055, respectively.} in Table~\ref{Table:nh_distribution} and in Fig.~\ref{fig:nh_compare}.

\begingroup
\renewcommand*{\arraystretch}{1.5}
\begin{table}
\caption{Intrinsic line-of-sight column density distribution derived from our new method, compared with previous results from observational works and from population synthesis models.}
\centering
\label{Table:nh_distribution}
\small
\begin{tabular}{ccccc}
       \hline
        \hline
&10$^{20}$--10$^{22}$&10$^{22}$--10$^{26}$&10$^{22}$--10$^{24}$&10$^{24}$--10$^{26}$\\
       \hline
Our work&33\%&67\%&30\%&37\%\\
        \hline
\end{tabular}
\begin{tabular}{c}
Observations\\
\end{tabular}

  \begin{tabular}{ccccc}
          \hline
	\hline
&10$^{20}$--10$^{22}$&10$^{22}$--10$^{25}$&10$^{22}$--10$^{24}$&10$^{24}$--10$^{25}$\\
       \hline
	Burlon+11&38\%&62\%&43\%&19\%\\
	\hline
	Ricci+15&22\%&78\%&46\%&32\%\\
	\hline
	\hline
\end{tabular}

\begin{tabular}{c}
population synthesis model\\
\end{tabular}

\begin{tabular}{ccccc}
       \hline
       \hline     
&10$^{20}$--10$^{22}$&10$^{22}$--10$^{26}$&10$^{22}$--10$^{24}$&10$^{24}$--10$^{26}$\\
       \hline
	Gilli+07&13\%&87\%&42\%&45\%\\
	\hline
	Ueda+14&33\%&67\%&34\%&33\%\\
	\hline
	Buchner+15&25\%&75\%&37\%&38\%\\
	\hline
	Ananna+19&9\%&91\%&41\%&50\%\\
	\hline
	\hline
\end{tabular}
\end{table}
\endgroup

When comparing to the \bat\ observed column density distributions \citep{Burlon11,Ricci15}, our method predicts a larger fraction of CT-AGN (see, Table~\ref{Table:nh_distribution}), which is likely due to the difficulty of \bat\ in detecting extremely heavily obscured AGN. Our predictions are in good agreement with the column density distribution used in the population synthesis model in \citet{Ueda14}, which are also based on a low-redshift \bat\ selected sample. It is worth noting that the column density distribution is luminosity (and redshift) dependent \citep[see, e.g.,][]{Buchner2015}. Different works adopt different samples of AGN with different luminosity and redshift ranges when deriving the column density distribution. In particular, our work provides a constraint on the line-of-sight column density distribution of AGN with the median 2--10\,keV intrinsic luminosity of $\left<\rm L_{int,2-10}\right>$ = 1.20 $\times$ 10$^{43}$\,erg\,s$^{-1}$ in the local Universe ($z\,<$0.15). The listed column density distributions of each work in Table~\ref{Table:nh_distribution} are the values at $z\sim$0. We note that the fraction of unobscured AGN derived in our work at $z\sim$0 is larger than those obtained in \citet{gilli07,Buchner2015,Tasnim_Ananna_2019}. Therefore, an overestimation of the CXB at soft X-ray band might be made if the fraction of unobscured AGN is consistent at different redshift, suggesting that the unobscured AGN fraction should significantly decrease moving towards higher redshifts, as discovered in observations \citep[e.g.,][]{Lanzuisi2018}.

\begin{figure} 
\centering
\includegraphics[width=.46\textwidth]{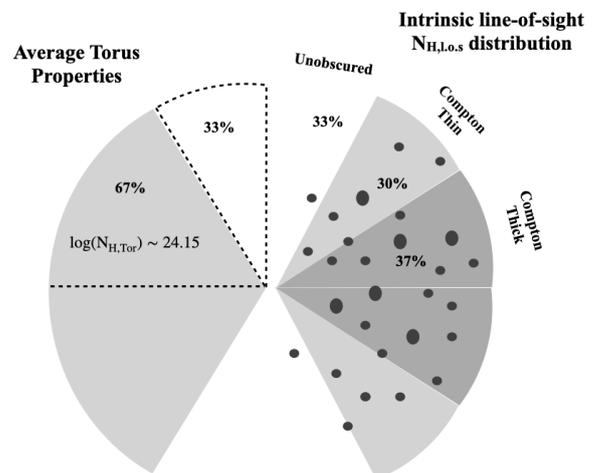}
\caption{The left part of the figure presents that the average torus properties of our unbiased Compton thin sample; The right part of the figure illustrates the intrinsic line-of-sight column density distribution derived from our new developed method.}
\label{fig:illustration}
\end{figure}

A schematic of the comparison of the average torus properties obtained from our unbiased sample (left side) and the intrinsic line-of-sight column density distribution derived from our method is illustrated in Fig.~\ref{fig:illustration}.

\section{Conclusions}
We study the obscuring torus of AGN by analyzing the broadband X-ray spectra of a large and unbiased sample of heavily obscured AGN in the local universe. We find that Compton thin and CT-AGN may possess similar tori, whose average column density is Compton thick (N$\rm_{H,tor,ave}\,\approx$\,1.4 $\times$ 10$^{24}$\,cm$^{-2}$), but they are observed through different (under-dense or over-dense) regions of the tori. The average torus covering factor of the Compton thin sample is $c\rm_{f}$ = 0.67, suggesting that the fraction of unobscured AGN is $\sim$33\%. We also find that the obscuring tori of most AGN are significantly inhomogeneous. Using the obtained properties of the obscuring torus and a Monte Carlo method, we calculate the intrinsic line-of-sight column density distribution of AGN in the nearby universe, which is in good agreement with the recent AGN population synthesis models. Our results may help understand the properties of the obscuring materials surrounding the SMBH in the local universe and help constrain the future population synthesis study of the CXB. In the future, new X-ray missions such as {\it Athena} and the proposed {\it Lynx} and {\it AXIS} facilities will detect a large sample of obscured AGN at high redshift \citep[see, e.g.,][]{marchesi2020}, which could help us further constrain the evolution of the AGN obscuration and SMBH growth over cosmic time.
%
%

\begin{acknowledgements}
X.Z. thanks the anonymous referee for their detailed and useful comments, which helped improve the paper significantly. X.Z., M.A. and S.M. acknowledge NASA funding under contract 80NSSC17K0635 and 80NSSC19K0531. X.Z. also acknowledges NASA funding under contract 80NSSC20K0043. This research has made use of the Palermo BAT Catalogue and database operated at INAF-IASF Palermo. We thank the \NuSTAR\ Operations, Software and Calibrations teams for support with these observations. This research has made use of data and/or software provided by the High Energy Astrophysics Science Archive Research Center (HEASARC), which is a service of the Astrophysics Science Division at NASA/GSFC and the High Energy Astrophysics Division of the Smithsonian Astrophysical Observatory. This work is based on observations obtained with \XMM, an ESA science mission with instruments and contributions directly funded by ESA Member States and NASA. The scientific results reported in this article are based in part on observations made by data obtained from the \cha\ Data Archive. This work made use of data supplied by the UK {\it Swift} Science Data Centre at the University of Leicester.
\end{acknowledgements}

\bibliographystyle{aa}
\bibliography{referencezxr}

\begin{appendix} 

\section{Data Reduction \& Spectral Analysis Results} \label{sec:data_reduction}
The information about the observations adopted when analyzing each source is listed in Table~\ref{Table:Sources1} and Table~\ref{Table:Sources2}. The redshift of each source is adopted from NED\footnote{\url{https://ned.ipac.caltech.edu}} and SIMBAD\footnote{\url{http://simbad.u-strasbg.fr/simbad/}}, except for 2MASX J00502684+8229000, which does not have redshift record. Thus, we let the redshift of this source free to vary when fitting its spectra, and we obtain a best-fit redshift of $z\sim$0.03817.

The \NuSTAR\ data of both FPMA and FPMB are calibrated, cleaned, and screened using the \texttt{nupipeline} script version 0.4.6 and calibration database (CALDB) version 20181030. The sources spectra, ancillary response files (ARF), and response matrix files (RMF) are obtained using the \texttt{nuproducts} script version 0.3.0. The sources spectra are extracted from a 75$^{\prime\prime}$ circular region, and the background spectra are extracted using a 75$^{\prime\prime}$ circular region near the source but avoiding contamination from the source.

The \XMM\ data from two MOS cameras \citep[][]{MOS} and one EPIC CCD cameras \citep[pn;][]{pn} on-board \XMM\ are utilized in the spectral analysis. The \XMM\ data are reduced using the Science Analysis System \citep[SAS;][]{SAS} version 17.0.0 following standard procedures. The flares are removed by visually inspecting the high energy light curve (10--12\,keV) when the count rates exceed 0.35\,cts/s for MOS and 0.4\,cts/s for pn. The source spectra are extracted from a circular region with a radius of 15$^{\prime\prime}$; the background spectra are extracted from a circle nearby the source with the same radius as the source spectra but avoiding contamination from sources.

\cha\ data, focused on the 0.3--7\,keV energy band, are used in our spectral analysis when the \XMM\ data are not available. We reduced the \cha\ data using the \cha's data analysis system, CIAO software package \citep{CIAO} version 4.11 and \cha\ CALDB version 4.8.2. The level=1 data are reprocessed using the CIAO \texttt{chandra\_repro} script. The source spectrum is extracted from a circular region with a radius of 5$^{\prime\prime}$; background spectrum is extracted from a circular region nearby the source with a radius of 10$^{\prime\prime}$. The CIAO \texttt{specextract} tool is used to extract both source and background spectra, ARF, and RMF files following standard procedures.

Given the smaller effective area in soft X-ray band (8 times smaller than \XMM\ pn and four times smaller than \cha\ at 3\,keV), \XRT\ data are used only when neither \XMM\ nor \cha\ data are available. The spectra are obtained from the online {\it Swift} product generator\footnote{\url{http://www.swift.ac.uk/user_ objects/}} \citep[see also][]{Evans09}. In the case of variability, we did not stack multiple observations of the same source. 
The \XRT\ spectra are rebinned with a minimum of 5 counts per bin due to the limited number of counts, using the HEAsoft task \texttt{grppha}, while the \NuSTAR, \XMM\ and \cha\ spectra are rebinned with a minimum of 20 counts per bin.

\begingroup
\renewcommand*{\arraystretch}{1.6}
\begin{table*}
\centering
\scriptsize
\caption{93 BAT selected sources with archival \NuSTAR\ observations.Col 2: redshift of the source. Col 3--4: \NuSTAR\ observation date and exposure time. Col 5--6: \XMM\ observation date and exposure time. Col 7--8: \cha\ observation date and exposure time. Col 9--10: XRT observation date and exposure time.
$^a$) No redshift record is found for 2MASX J00502684+8229000 in either NED or SIMBAD, so we let the redshift free to vary when fitting the spectra. The best-fit redshift is $z\sim$0.03817.}
\label{Table:Sources1}
  \begin{tabular}{lcccccccccccc}
       \hline
       \hline     
       Source&$z$&\NuSTAR&\NuSTAR&XMM&XMM&\cha&\cha&XRT&XRT\\
       &&date&ks&date&ks&date&ks&date&ks\\
       \hline
	2MASX J00502684+8229000&N/A$^a$&2018 Aug 25&54&&&&&2018 Aug 25&7\\
	2MASX J01073963-1139117&0.04746&2016 Sep 16&42&&&2019 May 21&10\\
	2MASX J06411806+3249313&0.04700&2018 Dec 31&36&2006 Mar 11&17\\
	2MASX J09235371-3141305&0.04237&2014 Apr 20&42&&&&&2014 Apr 19&7\\
	2MASX J11140245+2023140&0.02615&2014 Jan 17&48&&&&&2014 Jan 17&6\\
	3C 105&0.08900&2016 Aug 21&42&2008 Feb 25&34\\
	3C 403&0.05900&2013 May 25&40&&&2002 Dec 07&49\\
	3C 445&0.05588&2016 May 15&40&2001 Dec 06&57\\
	3C 452&0.08110&2017 May 01&104&2008 Nov 30&194\\
	4C +29.30&0.06471&2013 Nov 08&42&2008 Apr 11&62\\
	4C +73.08&0.05810&2016 Dec 05&26&&&2009 July 11&29\\
	CASG218&0.05395&2016 Dec 31&40&&&2016 Dec 26&23\\
	Cen A&0.00183&2013 Aug 06&102&2013 Aug 07&29\\
	CGCG 427-028&0.03033&2014 Dec 10&52&2014 Dec 10&84\\
	ESO 21-4&0.00984&2015 May 05&42&&&&&2019 Dec 22&6\\
	ESO 103-35&0.01329&2017 Oct 15&88&2002 Mar 15&30\\
	ESO 119-8&0.02294&2016 Jun 27&52&&&2016 July 09&10\\
	ESO 121-28&0.04052&2014 Aug 08&44&&&&&2014 Aug 08&7\\
	ESO 231-26&0.06254&2015 Mar 13&48&&&&&2009 Jun 19&2\\
	ESO 263-13&0.03354&2015 Oct 13&46&2007 Jun 14&61\\
	ESO 317-41&0.01932&2018 May 27&62&2018 May 27&72\\
	ESO 383-18&0.01241&2016 Jan 20&213&2006 Jan 10&45\\
	ESO 439-G009&0.02389&2015 Dec 20&46&2015 Jun 07&34\\
	ESO 464-16&0.03635&2016 Apr 13&44&2016 Apr 11&55\\
	ESO 500-34&0.01222&2017 Dec 15&36&&&2016 Feb 08&10\\
	ESO 505-30&0.03971&2018 Jan 28&44&&&&&2018 Jan 28&7\\
	ESO 553-43&0.02776&2016 Aug 16&43&&&&&2010 May 16&10\\
	Fairall 272&0.02182&2014 Jan 10&48&2007 Oct 14&30\\
	IC 751&0.03120&2014 Nov 28&52&2014 Nov 28&64\\
	IC 1657&0.01195&2017 Jan 15&90&&&&&2008 Nov 03&2\\
	IC 4709&0.01690&2013 July 24&40&&&&&2006 Oct 28&2\\
	IC 4518A&0.01626&2013 Aug 02&16&2006 Aug 07&22\\
	IC 5063&0.01135&2013 July 08&36&&&2007 Jun 15&34\\
	IRAS 05581+0006&0.11470&2015 Dec 12&42&&&&&2010 May 07&10\\
	IRAS 16288+3929&0.03056&2012 Sep 19&32&2011 Sep 08&78\\
	IRAS 20247-7542&0.11394&2019 Feb 18&70&&&&&2010 July 02&7\\
	LEDA 15023&0.04500&2012 July 24&12&&&2012 Oct 20&19\\
	LEDA 2265450&0.02678&2014 Dec 23&40&&&&&2006 Jan 04&6\\
	LEDA 259433&0.09001&2019 Mar 01&44&&&&&2009 Dec 31&7\\
	LEDA 2816387&0.10756&2016 May 06&53&2017 Nov 20&50\\
	LEDA 46599&0.03194&2016 Jan 09&42&&&&&2016 Jan 09&6\\
	LEDA 511869&0.07661&2016 May 25&44&2018 Feb 09&41\\
	LEDA 549777&0.06100&2014 Jan 21&44&&&&&2006 Jun 23&12\\
	MCG -01-05-047&0.01720&2012 Nov 30&26&2009 July 24&30\\
	MCG -02-09-040&0.01499&2015 Feb 25&44&2015 Feb 25&26\\
	MCG -02-15-004&0.02897&2018 Nov 13&40&&&&&2010 May 05&6\\
	MCG -03-34-064&0.01654&2016 Jan 17&157&2016 Jan 17&372\\
	MCG +11-11-32&0.03625&2016 Feb 18&8&&&&&2015 Dec 14&7\\
       \hline
       \hline
	\vspace{0.02cm}
\end{tabular}
\end{table*}
\endgroup

\begingroup
\renewcommand*{\arraystretch}{1.6}
\begin{table*}
\centering
\caption{(Continued)}
\label{Table:Sources2}
\scriptsize
  \begin{tabular}{lccccccccccc}
       \hline
       \hline     
       Source&$z$&\NuSTAR&\NuSTAR&XMM&XMM&\cha&\cha&XRT&XRT\\
       &&date&ks&date&ks&date&ks&date&ks\\
       \hline
	Mrk 3&0.01351&2015 Apr 08&50&2015 Apr 08&8\\
	Mrk 18&0.01109&2013 Dec 15&40&2006 Mar 23&36\\
	Mrk 273&0.03778&2013 Nov 04&140&2013 Nov 04&19\\
	Mrk 348&0.01503&2015 Oct 28&42&2002 July 18&76\\
	Mrk 417&0.03276&2017 Feb 20&42&2006 Jun 15&24\\
	Mrk 477&0.03773&2014 May 15&36&2010 July 21&30\\
	...&...&2014 May 24&34&...&...\\
	Mrk 1210&0.01350&2012 Oct 05&30&2001 May 05&18\\
	Mrk 1498&0.05470&2015 May 11&48&2007 Jun 23&21\\
	NGC 454E&0.01213&2016 Feb 14&48&2009 Nov 05&76\\
	NGC 612&0.02977&2012 Sep 14&30&2006 Jun 26&71\\
	NGC 788&0.01360&2013 Jan 28&30&2010 Jan 15&71\\
	NGC 835&0.01359&2015 Sep 13&42&2000 Jan 23&91\\
	NGC 1142&0.02885&2017 Oct 14&41&2006 Jan 28&32\\
	NGC 1229&0.03629&2013 July 05&50&&&&&2010 Oct 19&7\\
	NGC 2655&0.00467&2016 Nov 10&32&2005 Sep 04&16\\
	NGC 3281&0.01067&2016 Jan 22&46&2011 Jan 05&64\\
	NGC 4102&0.00282&2015 Nov 19&42&&&2016 Jun 17&29\\
	NGC 4258&0.00149&2016 Jan 10&208&2006 Nov 17&142\\
	NGC 4388&0.00842&2013 Dec 27&42&&&2008 Apr 16&171\\
	NGC 4500&0.01038&2015 Aug 30&42&&&&&2015 Aug 30&7\\
	NGC 4507&0.01180&2015 Jun 10&68&2010 Aug 03&29\\
	NGC 4785&0.01227&2014 Aug 20&98&2014 Aug 20&72\\
	NGC 4939&0.01037&2017 Feb 17&44&&&2010 Jun 09&14\\
	NGC 4941&0.00370&2016 Jan 19&42&&&2011 Mar 07&5\\
	NGC 4992&0.02514&2015 Jan 27&46&2006 Jun 27&45\\
	NGC 5283&0.01040&2018 Nov 17&66&&&2003 Nov 24&9\\
	NGC 5728&0.00935&2013 Jan 02&48&&&2003 Jun 27&19\\
	NGC 5899&0.00855&2014 Apr 08&48&2011 Feb 13&62\\
	NGC 5972&0.02974&2018 Mar 20&83&&&2017 Dec 18&24\\
	...&...&...&...&&&2017 Dec 19&24\\
	NGC 6300&0.00370&2016 Aug 24&48&2001 Mar 02&128\\
	NGC 6232&0.01480&2013 Aug 17&38&&&&&2011 Apr 29&3\\
	NGC 7319&0.02251&2017 Sep 27&84&2001 Dec 07&121\\
	PKS 2356-61&0.09631&2014 Aug 10&46&2011 Oct 19&22&2010 Mar 04&20\\
	SDSS J135429.05+132757.2&0.06332&2019 Mar 14&39&&&2014 Jun 25&9\\
	SDSS J165315.05+234942.9&0.10342&2018 Jan 19&56&2011 Mar 09&35\\
	UGC 3157&0.01541&2014 Mar 18&40&&&&&2010 Oct 22&6\\
	UGC 3752&0.01569&2013 Dec 03&48&&&&&2013 Dec 03&7\\
	UGC 3995B&0.01597&2014 Nov 08&46&&&2014 Feb 07&11\\
	UGC 4211&0.03443&2017 Mar 11&40&&&2019 Feb 08&10\\
	VII Zw 73&0.04133&2012 Nov 08&34&2011 Mar 04&27\\
	Was 49b&0.06400&2014 July 15&40&&&2014 July 15&6\\
	WISE J144850.99-400845.6&0.12323&2019 Mar 12&40&2014 Feb 16&67\\
	Z319-7&0.04400&2014 May 01&30&&&&&2014 May 01&7\\
	Z333-49&0.03358&2015 Jan 15&50&&&&&2009 Jan 14&10\\
	Z367-9&0.02392&2014 Dec 21&60&&&&&2007 May 30&10\\
       \hline
       \hline
       \vspace{0.02cm}
\end{tabular}
\end{table*}
\endgroup

\begingroup
\renewcommand*{\arraystretch}{1.45}
\begin{table*}
\centering
  \scriptsize
\caption{\normalsize Best-fit results of the sources with \NuSTAR\ (3-78\,keV) data \& \XMM\ (1-10\,keV) data.\\
Col 2: Reduced $\chi^2$, i.e., $\chi^2$/ degree of freedom (d.o.f.). Col 4: Logarithm of line-of-sight column density in cm$^{-2}$. If line-of-sight variability is found, the line-of-sight column density measured using \NuSTAR\ data are reported. Col 5: Logarithm of average column density of the torus in cm$^{-2}$. Col 6: Cosine of the inclination angle. cos($\theta$) = 1 is face-on and cos($\theta$) = 0.05 is edge-on. Col 7: Covering factor of the torus. Col 8: normalization of the main cut-off power-law component at 1\,keV in 10$^{-2}$ photons keV$^{-1}$\,cm$^{-2}$\,s$^{-1}$. Col 9: Logarithm of 2--10\,keV flux of \NuSTAR\ observation in $10^{-12}$\,erg\,cm$^{-2}$\,s$^{-1}$. Col 10: Logarithm of 2--10\,keV intrinsic luminosity of \NuSTAR\ observation in erg\,s$^{-1}$. The 1\,$\sigma$ confidence level error is reported here. {\it u} is unconstrained when fitting the spectra. Sources labeled with asterisks are those fitted with only the line-of-sight component and the scattered component.}
\label{Table:resultsXMM}
  \begin{tabular}{lccccccccccccccc}
       \hline
       \hline     
         Source&$\chi^2$/d.o.f.&$\Gamma$&N$\rm _{H,l.o.s}$&N$\rm _{H,tor}$&cos($\theta\rm _{inc}$)&$c_{\rm f}$&norm&Flux$_{2-10}$&L$_{\rm int,2-10}$\\
       \hline
       2MASX J06411806+3249313&382/371&1.60$_{-0.08}^{+0.09}$&23.09$_{-0.10}^{+0.08}$&22.86$_{-0.34}^{+0.89}$&0.95$_{-u}^{+u}$&0.88$_{-0.76}^{+u}$&0.16$_{-0.03}^{+0.04}$&--11.27$_{-0.08}^{+0.04}$&43.70$_{-0.05}^{+0.05}$\\
      \hline
     3C 105$^*$&205/232&1.53$_{-0.13}^{+0.13}$&23.65$_{-0.03}^{+0.03}$&-&-&-&0.25$_{-0.07}^{+0.10}$&--12.03$_{-0.05}^{+0.02}$&43.97$_{-0.07}^{+0.07}$\\ 
      \hline
       3C 445&885/913&1.56$_{-0.14}^{+0.05}$&23.08$_{-0.07}^{+0.05}$&24.00$_{-0.32}^{+0.11}$&0.95$_{-0.08}^{+u}$&0.80$_{-0.20}^{+0.12}$&0.22$_{-0.04}^{+0.02}$&--11.27$_{-0.08}^{+0.02}$&43.76$_{-0.01}^{+0.01}$\\
      \hline
       3C 452&1206/1177&1.49$_{-0.02}^{+0.02}$&23.59$_{-0.01}^{+0.03}$&22.96$_{-0.08}^{+0.04}$&0.95$_{-0.09}^{+u}$&1.00$_{-0.12}^{+u}$&0.26$_{-0.01}^{+0.02}$&--11.51$_{-0.02}^{+0.01}$&44.36$_{-0.01}^{+0.02}$\\  
       \hline
       4C +29.30&168/176&1.46$_{-u}^{+0.21}$&23.50$_{-0.06}^{+0.04}$&23.90$_{-0.09}^{+0.11}$&0.55$_{-u}^{+u}$&1.00$_{-0.57}^{+u}$&0.05$_{-0.01}^{+0.04}$&--12.00$_{-0.51}^{+0.01}$&43.35$_{-0.05}^{+0.13}$\\  
      \hline
      Cen A&5100/4781&1.81$_{-0.01}^{+0.01}$&23.06$_{-0.01}^{+0.01}$&23.04$_{-0.06}^{+0.26}$&0.35$_{-0.04}^{+0.15}$&0.30$_{-0.13}^{+0.04}$&9.95$_{-0.14}^{+0.07}$&--9.36$_{-0.01}^{+0.01}$&42.82$_{-0.01}^{+0.01}$\\  
      \hline
      CGCG 427-028&463/470&1.76$_{-0.09}^{+0.03}$&23.15$_{-0.04}^{+0.03}$&25.46$_{-1.12}^{+u}$&0.60$_{-0.40}^{+0.31}$&0.62$_{-0.42}^{+0.06}$&0.11$_{-0.02}^{+0.01}$&--11.66$_{-0.03}^{+0.02}$&42.99$_{-0.03}^{+0.03}$\\
      \hline
      ESO 103-35&2006/1907&1.70$_{-0.08}^{+0.10}$&23.25$_{-0.02}^{+0.02}$&24.36$_{-0.07}^{+0.08}$&0.51$_{-0.05}^{+0.16}$&0.65$_{-0.07}^{+0.14}$&1.73$_{-0.16}^{+0.34}$&--10.74$_{-0.06}^{+0.01}$&43.37$_{-0.02}^{+0.02}$\\
      \hline
      ESO 263-13&410/444&1.64$_{-0.12}^{+0.12}$&23.85$_{-0.03}^{+0.04}$&22.60$_{-0.12}^{+0.48}$&0.64$_{-u}^{+0.20}$&0.70$_{-0.42}^{+u}$&0.24$_{-0.01}^{+0.01}$&--11.87$_{-0.07}^{+0.01}$&43.50$_{-0.05}^{+0.04}$ \\ 
       \hline
     ESO 317-41&158/156&1.76$_{-0.21}^{+0.31}$&23.90$_{-0.08}^{+0.07}$&24.66$_{-0.50}^{+u}$&0.45$_{-u}^{+0.36}$&0.56$_{-0.26}^{+u}$&0.12$_{-0.07}^{+0.16}$&--12.36$_{-u}^{+0.17}$&42.60$_{-0.29}^{+0.18}$\\  
      \hline
      ESO 383-18&1367/1343&1.58$_{-0.05}^{+0.04}$&23.28$_{-0.02}^{+0.03}$&24.38$_{-0.10}^{+0.07}$&0.84$_{-0.15}^{+0.07}$&0.86$_{-0.13}^{+0.10}$&0.39$_{-0.04}^{+0.04}$&--11.34$_{-0.04}^{+0.02}$&42.54$_{-0.02}^{+0.03}$\\ 
      \hline
      ESO 439-G009&173/180&1.77$_{-0.19}^{+0.26}$&23.58$_{-0.09}^{+0.09}$&24.26$_{-0.15}^{+u}$&0.10$_{-u}^{+0.67}$&0.64$_{-0.38}^{+u}$&0.16$_{-0.07}^{+0.16}$&--11.93$_{-0.35}^{+0.17}$&42.89$_{-0.16}^{+0.14}$\\
      \hline
      ESO 464-16&137/131&1.46$_{-u}^{+0.30}$&23.79$_{-0.05}^{+0.18}$&23.96$_{-0.20}^{+0.27}$&0.74$_{-u}^{+0.11}$&0.91$_{-0.52}^{+u}$&0.04$_{-0.09}^{+0.05}$&--12.27$_{-u}^{+0.08}$&42.97$_{-0.10}^{+0.19}$\\   
      \hline
      Fairall 272&559/594&1.51$_{-0.08}^{+0.11}$&23.23$_{-0.05}^{+0.05}$&24.00$_{-0.31}^{+0.33}$&0.37$_{-u}^{+0.21}$&0.40$_{-0.15}^{+0.47}$&0.18$_{-0.04}^{+0.06}$&--11.19$_{-0.20}^{+0.06}$&43.10$_{-0.04}^{+0.06}$\\
      \hline
      IC 4518A&188/186&1.72$_{-0.18}^{+0.15}$&23.11$_{-0.16}^{+0.12}$&24.28$_{-0.30}^{+0.40}$&0.75$_{-u}^{+0.16}$&0.80$_{-0.37}^{+u}$&0.20$_{-0.06}^{+0.09}$&--11.37$_{-0.09}^{+0.03}$&42.75$_{-0.09}^{+0.06}$\\ 
      \hline
      IC 751&280/257&2.06$_{-0.14}^{+0.15}$&23.60$_{-0.03}^{+0.03}$&24.76$_{-0.30}^{+u}$&0.59$_{-u}^{+0.27}$&0.72$_{-0.24}^{+u}$&0.28$_{-0.10}^{+0.09}$&--11.93$_{-0.07}^{+0.10}$&43.20$_{-0.09}^{+0.06}$\\   
      \hline
      IRAS 16288+3929&187/196&1.72$_{-0.17}^{+0.12}$&23.88$_{-0.10}^{+0.06}$&23.42$_{-0.12}^{+0.69}$&0.26$_{-u}^{+0.47}$&0.41$_{-0.17}^{+0.44}$&0.23$_{-0.11}^{+0.11}$&--12.06$_{-0.07}^{+0.06}$&43.28$_{-0.19}^{+0.17}$\\   
      \hline
     LEDA 2816387$^*$&243/232&1.33$_{-0.12}^{+0.13}$&23.76$_{-0.07}^{+0.05}$&-&-&-&0.08$_{-0.03}^{+0.04}$&--12.00$_{-0.09}^{+0.01}$&44.24$_{-0.10}^{+0.11}$\\ 
           \hline
      LEDA 511869&157/155&1.49$_{-u}^{+0.18}$&23.88$_{-0.03}^{+0.12}$&24.00$_{-0.70}^{+u}$&0.13$_{-0.06}^{+0.13}$&0.10$_{-u}^{+0.26}$&0.22$_{-0.05}^{+0.23}$&--11.24$_{-0.61}^{+0.03}$&43.92$_{-0.06}^{+0.07}$\\  
      \hline
	MCG -01-05-047&251/261&1.89$_{-0.12}^{+0.10}$&23.33$_{-0.09}^{+0.09}$&24.46$_{-0.26}^{+u}$&0.89$_{-0.20}^{+0.05}$&0.89$_{-0.16}^{+0.07}$&0.30$_{-0.07}^{+0.08}$&--11.58$_{-0.06}^{+0.04}$&42.68$_{-0.04}^{+0.04}$\\ 
      \hline
     MCG -02-09-040&111/92&1.95$_{-0.15}^{+0.39}$&23.90$_{-0.14}^{+0.12}$&25.50$_{-1.14}^{+u}$&0.86$_{-0.52}^{+u}$&0.77$_{-0.43}^{+0.11}$&0.12$_{-0.07}^{+0.38}$&--12.31$_{-u}^{+0.01}$&42.26$_{-0.34}^{+0.24}$\\   
      \hline
      MCG -03-34-064&1920/1460&1.92$_{-0.02}^{+0.02}$&23.77$_{-0.01}^{+0.02}$&24.53$_{-0.05}^{+0.03}$&0.85$_{-0.00}^{+0.01}$&0.86$_{-0.00}^{+0.04}$&0.48$_{-0.03}^{+0.01}$&--11.68$_{-0.06}^{+0.02}$&42.92$_{-0.01}^{+0.01}$ \\ 
      \hline
       Mrk 3&1056/1073&1.48$_{-u}^{+0.11}$&23.94$_{-0.04}^{+0.06}$&23.30$_{-0.05}^{+0.24}$&0.47$_{-0.07}^{+0.16}$&0.50$_{-0.22}^{+0.06}$&1.42$_{-0.22}^{+0.54}$&--11.01$_{-1.05}^{+0.01}$&43.42$_{-0.04}^{+0.05}$\\ 
       \hline
      Mrk 18&212/204&1.70$_{-0.23}^{+0.15}$&23.13$_{-0.05}^{+0.06}$&25.10$_{-2.22}^{+u}$&0.95$_{-0.63}^{+u}$&0.25$_{-0.14}^{+0.10}$&0.07$_{-0.01}^{+0.02}$&--11.97$_{-0.23}^{+0.03}$&41.76$_{-0.05}^{+0.05}$\\  
      \hline
      Mrk 273&339/359&1.76$_{-0.20}^{+0.20}$&23.52$_{-0.09}^{+0.05}$&24.11$_{-0.22}^{+0.10}$&0.15$_{-u}^{+u}$&1.00$_{-0.44}^{+u}$&0.06$_{-0.03}^{+0.00}$&--12.09$_{-0.27}^{+0.01}$&42.90$_{-0.12}^{+0.14}$\\  
      \hline
	Mrk 348&2702/2583&1.66$_{-0.03}^{+0.03}$&22.92$_{-0.04}^{+0.03}$&24.48$_{-0.54}^{+0.48}$&0.21$_{-0.03}^{+0.02}$&0.17$_{-u}^{+0.08}$&1.33$_{-0.08}^{+0.09}$&--10.43$_{-0.01}^{+0.01}$&43.48$_{-0.08}^{+0.04}$ \\ 
      \hline
      Mrk 417&314/326&1.57$_{-0.09}^{+0.08}$&23.53$_{-0.07}^{+0.06}$&23.15$_{-0.25}^{+1.53}$&0.15$_{-u}^{+0.26}$&0.40$_{-u}^{+0.21}$&0.23$_{-0.05}^{+0.04}$&--11.51$_{-0.05}^{+0.01}$&43.46$_{-0.01}^{+0.01}$\\ 
            \hline
      Mrk 477&673/539&1.58$_{-0.07}^{+0.09}$&23.30$_{-0.07}^{+0.06}$&22.90$_{-0.20}^{+0.14}$&0.95$_{-u}^{+u}$&1.00$_{-0.22}^{+u}$&0.16$_{-0.03}^{+0.06}$&--11.47$_{-0.01}^{+0.01}$&43.36$_{-0.05}^{+0.05}$\\  
      \hline
      Mrk 1210&679/669&1.77$_{-0.03}^{+0.06}$&23.30$_{-0.02}^{+0.04}$&24.11$_{-0.08}^{+0.11}$&0.95$_{-u}^{+u}$&1.00$_{-0.07}^{+u}$&0.74$_{-0.11}^{+0.15}$&--11.12$_{-0.02}^{+0.01}$&42.92$_{-0.03}^{+0.03}$\\   
      \hline
     Mrk 1498&811/815&1.67$_{-0.04}^{+0.04}$&23.26$_{-0.03}^{+0.02}$&23.00$_{-0.70}^{+0.63}$&0.95$_{-0.62}^{+u}$&0.32$_{-u}^{+0.62}$&0.42$_{-0.05}^{+0.05}$&--11.10$_{-0.02}^{+0.01}$&44.12$_{-0.02}^{+0.02}$\\  
      \hline
      NGC 454E&344/310&1.81$_{-0.11}^{+0.11}$&23.86$_{-0.04}^{+0.04}$&24.75$_{-0.40}^{+0.68}$&0.43$_{-0.11}^{+0.16}$&0.40$_{-0.24}^{+0.26}$&0.16$_{-0.03}^{+0.06}$&--12.18$_{-0.15}^{+0.02}$&42.26$_{-0.04}^{+0.04}$\\ 
      \hline
	NGC 612&176/195&1.54$_{-0.08}^{+0.12}$&23.95$_{-0.04}^{+0.05}$&23.78$_{-0.29}^{+0.18}$&0.05$_{-u}^{+0.01}$&0.10$_{-u}^{+0.20}$&0.53$_{-0.02}^{+0.30}$&--11.96$_{-0.21}^{+0.02}$&43.56$_{-0.02}^{+0.14}$\\ 
	\hline
	NGC 788&303/290&1.56$_{-0.16}^{+0.14}$&23.79$_{-0.05}^{+0.04}$&24.50$_{-1.11}^{+u}$&0.26$_{-0.03}^{+0.08}$&0.20$_{-u}^{+0.25}$&0.45$_{-0.15}^{+0.18}$&--11.71$_{-0.25}^{+0.01}$&42.97$_{-0.10}^{+0.07}$\\  
	\hline
	NGC 835&104/124&1.40$_{-u}^{+0.25}$&23.46$_{-0.06}^{+0.09}$&23.22$_{-0.31}^{+0.16}$&0.05$_{-u}^{+0.34}$&0.40$_{-0.10}^{+0.20}$&0.04$_{-0.01}^{+0.01}$&--12.09$_{-0.58}^{+0.12}$&41.81$_{-0.08}^{+0.02}$\\
	\hline
	NGC 1142&225/255&1.48$_{-u}^{+0.16}$&24.20$_{-0.07}^{+0.08}$&23.49$_{-0.13}^{+0.15}$&0.05$_{-u}^{+0.28}$&0.45$_{-0.05}^{+0.14}$&0.36$_{-0.08}^{+0.17}$&--12.10$_{-0.16}^{+0.04}$&43.44$_{-0.11}^{+0.13}$\\ 
	\hline
	NGC 2655&123/120&2.12$_{-0.17}^{+0.23}$&23.51$_{-0.08}^{+0.08}$&23.65$_{-u}^{+u}$&0.15$_{-u}^{+0.64}$&0.39$_{-u}^{+u}$&0.22$_{-0.10}^{+0.02}$&--11.96$_{-2.54}^{+0.07}$&41.37$_{-0.16}^{+0.18}$\\  
      \hline
      NGC 3281&506/495&1.77$_{-0.09}^{+0.12}$&24.30$_{-0.06}^{+0.06}$&25.50$_{-0.08}^{+u}$&0.55$_{-0.09}^{+0.13}$&0.52$_{-0.13}^{+0.12}$&0.68$_{-0.19}^{+0.31}$&--11.72$_{-0.23}^{+0.10}$&42.98$_{-0.10}^{+0.11}$\\
      \hline
      NGC 4258&1577/1582&1.92$_{-0.05}^{+0.08}$&22.94$_{-0.02}^{+0.02}$&23.49$_{-0.15}^{+0.26}$&0.95$_{-0.05}^{+u}$&0.90$_{-0.05}^{+0.05}$&0.21$_{-0.02}^{+0.03}$&--11.66$_{-0.03}^{+0.01}$&40.23$_{-0.01}^{+0.02}$\\   
      \hline
      NGC 4507&1383/1355&1.63$_{-0.04}^{+0.05}$&23.87$_{-0.02}^{+0.02}$&22.90$_{-0.09}^{+0.16}$&0.51$_{-u}^{+0.11}$&0.66$_{-0.18}^{+0.12}$&1.49$_{-0.20}^{+0.32}$&--10.97$_{-0.01}^{+0.01}$&43.33$_{-0.04}^{+0.03}$\\   
      \hline
      NGC 4785&402/364&1.96$_{-0.15}^{+0.15}$&23.70$_{-0.04}^{+0.04}$&24.43$_{-0.37}^{+0.37}$&0.63$_{-0.05}^{+0.22}$&0.70$_{-0.11}^{+u}$&0.21$_{-0.08}^{+0.12}$&--12.04$_{-0.07}^{+0.12}$&42.32$_{-0.12}^{+0.07}$\\ 
      \hline
      NGC 4992&480/449&1.50$_{-0.05}^{+0.06}$&23.46$_{-0.03}^{+0.03}$&24.00$_{-0.07}^{+0.04}$&0.25$_{-u}^{+u}$&1.00$_{-0.22}^{+u}$&0.18$_{-0.03}^{+0.05}$&--11.54$_{-0.14}^{+0.01}$&43.10$_{-0.06}^{+0.04}$\\   
      \hline
     NGC 5899&820/866&1.99$_{-0.02}^{+0.03}$&23.08$_{-0.01}^{+0.01}$&24.79$_{-0.15}^{+0.13}$&0.95$_{-0.02}^{+u}$&0.95$_{-0.01}^{+0.01}$&0.43$_{-0.01}^{+0.01}$&--11.27$_{-0.02}^{+0.01}$&42.26$_{-0.01}^{+0.01}$\\  
      \hline
     NGC 6300&863/952&1.90$_{-0.08}^{+0.05}$&23.21$_{-0.05}^{+0.03}$&24.40$_{-0.10}^{+0.16}$&0.73$_{-0.14}^{+0.13}$&0.81$_{-0.14}^{+0.13}$&0.11$_{-0.02}^{+0.02}$&--10.81$_{-0.01}^{+0.03}$&42.08$_{-0.03}^{+0.02}$ \\  
      \hline
     NGC 7319&387/346&1.71$_{-0.21}^{+0.09}$&23.81$_{-0.05}^{+0.04}$&25.50$_{-0.38}^{+u}$&0.69$_{-0.08}^{+0.20}$&0.70$_{-0.26}^{+0.05}$&0.27$_{-0.12}^{+0.10}$&--12.39$_{-0.15}^{+0.08}$&42.56$_{-0.10}^{+0.08}$\\  
      \hline
      PKS 2356-61&296/294&1.90$_{-0.09}^{+0.04}$&23.23$_{-0.03}^{+0.03}$&25.50$_{-1.26}^{+u}$&0.76$_{-0.08}^{+0.17}$&0.76$_{-0.41}^{+0.07}$&0.21$_{-0.05}^{+0.02}$&--11.57$_{-0.06}^{+0.01}$&44.26$_{-0.03}^{+0.03}$\\   
      \hline
     SDSS J165315+234942&307/350&1.50$_{-0.06}^{+0.09}$&23.20$_{-0.05}^{+0.06}$&22.95$_{-0.65}^{+1.33}$&0.95$_{-0.80}^{+u}$&0.70$_{-u}^{+u}$&0.08$_{-0.01}^{+0.02}$&--11.70$_{-0.07}^{+0.03}$&44.01$_{-0.05}^{+0.04}$\\ 
      \hline
     VIIZw73$^*$&93/96&1.78$_{-0.26}^{+0.27}$&23.81$_{-0.09}^{+0.08}$&-&-&-&0.21$_{-0.12}^{+0.28}$&--12.17$_{-0.31}^{+0.05}$&43.49$_{-0.19}^{+0.20}$\\  
      \hline
     WISE J144850.99-400845.6$^*$&1247/1099&1.60$_{-0.04}^{+0.03}$&22.45$_{-0.03}^{+0.03}$&-&-&-&0.12$_{-0.01}^{+0.01}$&--11.28$_{-0.01}^{+0.01}$&44.74$_{-0.01}^{+0.01}$\\ 
      \hline
\end{tabular}
\vspace{0.2 cm}
\end{table*}
\endgroup

\begingroup
\renewcommand*{\arraystretch}{1.55}
\begin{table*}
\scriptsize
\centering
\caption{\normalsize Best-fit results of the sources with \NuSTAR\ (3-78\,keV) data \& \cha\ (1-7\,keV) data}
\label{Table:resultsCha}
  \begin{tabular}{lccccccccccccccc}
       \hline
       \hline     
         Source&$\chi^2$/d.o.f.&$\Gamma$&N$\rm _{H,l.o.s}$&N$\rm _{H,tor}$&cos($\theta\rm _{inc}$)&$c_{\rm f}$&norm&Flux$_{2-10}$&L$_{\rm int,2-10}$\\
       \hline
	2MASX J01073963-1139117&10/18&1.82$_{-0.20}^{+0.50}$&24.48$_{-0.05}^{+u}$&23.47$_{-0.38}^{+0.29}$&0.05$_{-u}^{+u}$&0.10$_{-u}^{+u}$&1.13$_{-0.12}^{+4.69}$&--12.33$_{-u}^{+0.15}$&44.30$_{-0.93}^{+1.44}$\\   
      \hline
	3C 403&358/305&2.11$_{-0.11}^{+0.09}$&23.65$_{-0.03}^{+0.03}$&24.76$_{-0.38}^{+u}$&0.56$_{-0.15}^{+0.25}$&0.60$_{-0.14}^{+0.28}$&0.67$_{-0.18}^{+0.19}$&--11.64$_{-0.03}^{+0.05}$&44.09$_{-0.05}^{+0.05}$\\   
      \hline
      4C +73.08&139/135&1.40$_{-u}^{+0.12}$&23.53$_{-0.12}^{+0.04}$&22.93$_{-0.47}^{+0.21}$&0.05$_{-u}^{+u}$&0.70$_{-0.31}^{+u}$&0.08$_{-0.01}^{+0.00}$&--11.81$_{-0.31}^{+0.10}$&43.60$_{-0.05}^{+0.02}$\\ 
      \hline
       CASG 218&201/209&1.60$_{-0.14}^{+0.19}$&23.41$_{-0.08}^{+0.04}$&24.13$_{-1.48}^{+u}$&0.35$_{-u}^{+u}$&0.20$_{-u}^{+0.71}$&0.11$_{-0.04}^{+0.06}$&--11.73$_{-0.15}^{+0.04}$&43.55$_{-0.10}^{+0.08}$\\   
      \hline
      ESO 119-8$^*$&91/106&1.66$_{-0.17}^{+0.18}$&23.17$_{-0.09}^{+0.09}$&-&-&-&0.03$_{-0.01}^{+0.02}$&--12.19$_{-0.07}^{+0.03}$&42.25$_{-0.07}^{+0.08}$\\  
      \hline
       ESO 500-34&115/120&2.34$_{-0.43}^{+0.25}$&23.67$_{-0.05}^{+0.09}$&24.17$_{-0.45}^{+u}$&0.15$_{-u}^{+0.71}$&0.88$_{-0.68}^{+u}$&0.62$_{-0.42}^{+0.72}$&--11.92$_{-0.56}^{+0.15}$&42.52$_{-0.27}^{+0.17}$\\ 
      \hline
       IC 5063&732/742&1.77$_{-0.07}^{+0.08}$&23.36$_{-0.02}^{+0.02}$&24.31$_{-0.21}^{+0.24}$&0.48$_{-0.11}^{+0.16}$&0.53$_{-0.15}^{+0.22}$&0.71$_{-0.12}^{+0.14}$&--11.06$_{-0.02}^{+0.02}$&42.87$_{-0.03}^{+0.03}$\\  
      \hline
      LEDA 15023&78/77&1.91$_{-0.25}^{+0.25}$&23.97$_{-0.18}^{+0.08}$&23.18$_{-0.28}^{+0.18}$&0.05$_{-u}^{+0.12}$&0.18$_{-u}^{+0.12}$&0.88$_{-0.51}^{+1.23}$&--11.86$_{-0.52}^{+0.03}$&44.09$_{-0.11}^{+0.10}$\\   
      \hline
      NGC 4102&297/300&1.74$_{-0.08}^{+0.12}$&23.87$_{-0.03}^{+0.04}$&25.15$_{-0.68}^{+u}$&0.53$_{-0.24}^{+0.16}$&0.61$_{-0.22}^{+0.31}$&0.46$_{-0.12}^{+0.22}$&--11.84$_{-0.04}^{+0.31}$&41.49$_{-0.06}^{+0.09}$\\  
      \hline
       NGC 4388&953/871&1.64$_{-0.08}^{+0.05}$&23.43$_{-0.02}^{+0.02}$&24.00$_{-0.20}^{+0.07}$&0.60$_{-0.04}^{+0.15}$&0.60$_{-0.05}^{+0.16}$&0.49$_{-0.09}^{+0.06}$&--11.14$_{-0.02}^{+0.01}$&42.52$_{-0.03}^{+0.03}$\\ 
      \hline
      NGC 4939&228/234&1.69$_{-0.15}^{+0.19}$&23.82$_{-0.08}^{+0.09}$&23.02$_{-0.14}^{+0.13}$&0.05$_{-u}^{+0.45}$&0.57$_{-0.14}^{+0.13}$&0.31$_{-0.13}^{+0.30}$&--11.80$_{-0.15}^{+0.01}$&42.49$_{-0.13}^{+0.16}$\\
      \hline
      NGC 4941&106/111&1.40$_{-u}^{+0.31}$&23.98$_{-0.20}^{+0.30}$&24.50$_{-0.68}^{+0.44}$&0.65$_{-0.20}^{+u}$&0.40$_{-0.23}^{+0.43}$&0.06$_{-0.03}^{+0.01}$&--12.28$_{-u}^{+0.06}$&41.04$_{-0.46}^{+0.43}$\\
      \hline
      NGC 5283&343/302&1.81$_{-0.17}^{+0.16}$&23.04$_{-0.09}^{+0.11}$&24.39$_{-0.80}^{+u}$&0.60$_{-0.19}^{+u}$&0.50$_{-0.32}^{+u}$&0.13$_{-0.04}^{+0.05}$&--11.62$_{-0.11}^{+0.01}$&42.02$_{-0.06}^{+0.06}$\\ 
      \hline
      NGC 5728&417/446&1.80$_{-0.10}^{+0.02}$&24.09$_{-0.01}^{+0.03}$&25.21$_{-0.61}^{+u}$&0.52$_{-0.01}^{+0.21}$&0.59$_{-0.11}^{+0.00}$&1.80$_{-0.03}^{+0.92}$&--11.78$_{-0.15}^{+0.52}$&43.08$_{-0.04}^{+0.09}$\\   
      \hline
      NGC 5972&177/188&1.75$_{-0.21}^{+0.36}$&23.79$_{-0.07}^{+0.13}$&24.34$_{-0.18}^{+u}$&0.85$_{-0.33}^{+0.03}$&0.90$_{-0.32}^{+u}$&0.08$_{-0.03}^{+0.22}$&--12.35$_{-u}^{+0.10}$&42.79$_{-0.09}^{+0.07}$\\   
      \hline
      SDSS J135429.05+132757.2$^*$&113/149&1.61$_{-0.15}^{+0.16}$&23.45$_{-0.11}^{+0.09}$&-&-&-&0.11$_{-0.04}^{+0.07}$&--11.83$_{-0.07}^{+0.01}$&43.67$_{-0.10}^{+0.11}$\\  
      \hline
      UGC 3995B&234/258&1.61$_{-0.09}^{+0.05}$&23.48$_{-0.02}^{+0.04}$&24.25$_{-0.16}^{+0.08}$&0.87$_{-u}^{+0.05}$&0.94$_{-0.18}^{+u}$&0.15$_{-0.03}^{+0.07}$&--11.68$_{-0.19}^{+0.02}$&42.58$_{-0.15}^{+0.03}$\\  
      \hline
      UGC 4211&295/304&1.77$_{-0.09}^{+0.11}$&22.95$_{-0.05}^{+0.05}$&23.43$_{-0.58}^{+0.41}$&0.05$_{-u}^{+u}$&0.70$_{-0.60}^{+u}$&0.15$_{-0.03}^{+0.07}$&--11.38$_{-0.13}^{+0.01}$&43.28$_{-0.05}^{+0.04}$\\ 
      \hline
       Was49b$^*$&215/171&1.49$_{-0.12}^{+0.13}$&23.30$_{-0.10}^{+0.11}$&-&-&-&0.08$_{-0.02}^{+0.04}$&--11.69$_{-0.06}^{+0.07}$&43.65$_{-0.09}^{+0.10}$\\   
      \hline

\end{tabular}
\vspace{0.3 cm}
\end{table*}
\endgroup

\begingroup
\renewcommand*{\arraystretch}{1.55}
\begin{table*}
\scriptsize
\centering
\caption{\normalsize Best-fit results of the sources with \NuSTAR\ (3-78\,keV) data \& \XRT\ (1-10\,keV) data.\\
Col 2: Reduced C statistic.}
\label{Table:resultsXRT}
  \begin{tabular}{lccccccccccccccc}
       \hline
       \hline     
         Source&Cstat/d.o.f.&$\Gamma$&N$\rm _{H,l.o.s}$&N$\rm _{H,tor}$&cos($\theta\rm _{inc}$)&$c_{\rm f}$&norm&Flux$_{2-10}$&L$_{\rm int,2-10}$\\
       \hline
	2MASX J00502684+8229000&128/160&2.02$_{-0.22}^{+0.21}$&23.21$_{-0.17}^{+0.19}$&24.15$_{-0.37}^{+0.17}$&0.95$_{-u}^{+u}$&0.98$_{-0.17}^{+u}$&0.17$_{-0.07}^{+0.15}$&--11.72$_{-0.19}^{+0.05}$&43.13$_{-0.27}^{+0.48}$\\ 
      \hline
	2MASX J09235371-3141305$^*$&183/148&1.37$_{-0.13}^{+0.14}$&23.67$_{-0.08}^{+0.07}$&-&-&-&0.13$_{-0.05}^{+0.08}$&--11.81$_{-0.06}^{+0.02}$&43.56$_{-0.11}^{+0.11}$\\ 
      \hline
	2MASX J11140245+2023140$^*$&106/109&1.53$_{-0.25}^{+0.27}$&23.70$_{-0.13}^{+0.11}$&-&-&-&0.08$_{-0.05}^{+0.12}$&--12.18$_{-0.68}^{+0.01}$&42.83$_{-0.19}^{+0.21}$ \\ 
      \hline
       ESO 21-4$^*$&215/180&1.57$_{-0.14}^{+0.15}$&23.43$_{-0.10}^{+0.10}$&-&-&-&0.12$_{-0.04}^{+0.08}$&--11.75$_{-0.05}^{+0.02}$&42.11$_{-0.10}^{+0.11}$\\
      \hline
       ESO 121-28&289/303&2.07$_{-0.08}^{+0.08}$&23.40$_{-0.08}^{+0.05}$&24.25$_{-0.19}^{+0.04}$&0.35$_{-u}^{+u}$&1.00$_{-0.29}^{+u}$&0.53$_{-0.20}^{+0.15}$&--11.47$_{-0.08}^{+0.02}$&43.67$_{-0.10}^{+0.05}$\\  
      \hline
       ESO 231-26&296/309&1.74$_{-0.14}^{+0.18}$&23.37$_{-0.07}^{+0.08}$&24.52$_{-0.49}^{+u}$&0.26$_{-u}^{+0.34}$&0.40$_{-u}^{+0.37}$&0.27$_{-0.12}^{+0.15}$&--11.50$_{-0.06}^{+0.04}$&43.96$_{-0.10}^{+0.09}$\\  
      \hline
       ESO 505-30&292/299&1.84$_{-0.14}^{+0.05}$&23.10$_{-0.09}^{+0.08}$&24.20$_{-0.33}^{+0.20}$&0.05$_{-u}^{+0.53}$&0.50$_{-0.19}^{+0.43}$&0.19$_{-0.04}^{+0.04}$&--11.41$_{-0.05}^{+0.05}$&43.48$_{-0.06}^{+0.06}$\\  
      \hline
      ESO 553-43$^*$&488/493&1.63$_{-0.05}^{+0.05}$&23.18$_{-0.04}^{+0.06}$&-&-&-&0.34$_{-0.05}^{+0.05}$&--11.19$_{-0.01}^{+0.01}$&43.42$_{-0.03}^{+0.03}$\\   
      \hline
	IC 1657&161/143&1.58$_{-u}^{+0.36}$&23.53$_{-0.21}^{+0.21}$&24.11$_{-1.09}^{+u}$&0.67$_{-0.36}^{+0.24}$&0.70$_{-0.58}^{+u}$&0.07$_{-0.03}^{+0.06}$&--11.94$_{-u}^{+0.04}$&42.00$_{-0.19}^{+0.28}$\\   
      \hline
       IC 4709&353/326&1.81$_{-0.15}^{+0.16}$&23.38$_{-0.12}^{+0.02}$&24.16$_{-0.22}^{+0.08}$&0.95$_{-u}^{+u}$&0.98$_{-0.06}^{+u}$&0.32$_{-0.11}^{+0.21}$&--11.42$_{-0.09}^{+0.04}$&42.77$_{-0.05}^{+0.07}$ \\  
      \hline
       IRAS 05581+0006&201/199&1.47$_{-u}^{+0.22}$&23.32$_{-0.12}^{+0.11}$&24.00$_{-0.32}^{+u}$&0.35$_{-u}^{+0.44}$&0.50$_{-u}^{+u}$&0.08$_{-0.02}^{+0.05}$&--11.72$_{-0.51}^{+0.04}$&44.15$_{-0.08}^{+0.12}$\\  
      \hline
       IRAS 20247-7542&446/401&1.62$_{-0.11}^{+0.07}$&23.02$_{-0.08}^{+0.07}$&24.03$_{-0.16}^{+0.17}$&0.90$_{-0.41}^{+u}$&0.35$_{-u}^{+0.54}$&0.11$_{-0.02}^{+0.02}$&--11.49$_{-0.04}^{+0.05}$&44.23$_{-0.05}^{+0.04}$\\  
      \hline
      LEDA 2265450$^*$&222/224&1.66$_{-0.11}^{+0.11}$&23.46$_{-0.06}^{+0.06}$&-&-&-&0.22$_{-0.06}^{+0.09}$&--11.62$_{-0.03}^{+0.01}$&43.18$_{-0.02}^{+0.03}$\\  
      \hline
      LEDA 46599&299/263&1.88$_{-0.26}^{+0.21}$&23.40$_{-0.12}^{+0.08}$&24.82$_{-2.12}^{+u}$&0.91$_{-0.38}^{+u}$&0.43$_{-u}^{+0.32}$&0.26$_{-0.07}^{+0.10}$&--11.54$_{-0.11}^{+0.01}$&43.35$_{-0.11}^{+0.11}$\\  
      \hline
       LEDA 259433&190/196&1.80$_{-0.01}^{+0.18}$&23.07$_{-0.21}^{+0.15}$&24.10$_{-0.30}^{+0.25}$&0.05$_{-u}^{+u}$&1.00$_{-0.17}^{+u}$&0.12$_{-0.05}^{+0.08}$&--11.65$_{-0.17}^{+0.01}$&43.93$_{-0.08}^{+0.09}$\\  
      \hline
      LEDA 549777$^*$&196/187&1.58$_{-0.10}^{+0.10}$&23.11$_{-0.11}^{+0.07}$&-&-&-&0.07$_{-0.01}^{+0.03}$&--11.75$_{-0.03}^{+0.01}$&43.39$_{-0.05}^{+0.06}$\\   
      \hline
       MCG+11-11-32&117/133&1.88$_{-0.10}^{+0.09}$&23.11$_{-0.07}^{+0.09}$&24.34$_{-0.19}^{+0.25}$&0.65$_{-u}^{+u}$&1.00$_{-0.32}^{+u}$&0.50$_{-0.12}^{+0.16}$&--11.16$_{-0.08}^{+0.03}$&43.67$_{-0.07}^{+0.04}$\\   
      \hline
      MCG-02-15-004&127/120&2.14$_{-0.07}^{+0.42}$&23.74$_{-0.11}^{+0.05}$&24.25$_{-0.22}^{+0.17}$&0.87$_{-u}^{+u}$&0.98$_{-0.50}^{+u}$&0.46$_{-0.02}^{+0.08}$&--11.77$_{-u}^{+0.09}$&43.30$_{-0.09}^{+0.15}$\\
      \hline
      NGC 1229&115/135&1.45$_{-u}^{+0.34}$&23.32$_{-0.09}^{+0.14}$&24.31$_{-0.28}^{+u}$&0.86$_{-0.37}^{+u}$&0.80$_{-0.53}^{+u}$&0.03$_{-0.00}^{+0.08}$&--12.11$_{-u}^{+0.03}$&42.77$_{-0.09}^{+0.20}$\\   
      \hline
       NGC 4500&40/30&1.54$_{-u}^{+0.58}$&23.36$_{-0.46}^{+0.21}$&23.52$_{-u}^{+0.78}$&0.95$_{-u}^{+u}$&0.71$_{-u}^{+u}$&0.01$_{-0.01}^{+0.02}$&--12.41$_{-u}^{+0.03}$&41.24$_{-0.18}^{+0.58}$\\   
      \hline
      NGC 6232&14/14&1.40$_{-u}^{+0.59}$&23.72$_{-0.08}^{+0.51}$&24.75$_{-0.81}^{+u}$&0.55$_{-u}^{+u}$&0.90$_{-0.74}^{+u}$&0.01$_{-0.01}^{+0.00}$&--12.68$_{-u}^{+0.69}$&41.56$_{-0.60}^{+0.09}$\\  
      \hline
       UGC 3157&121/134&2.21$_{-0.26}^{+0.34}$&23.85$_{-0.04}^{+0.07}$&24.84$_{-0.29}^{+0.10}$&0.91$_{-0.06}^{+u}$&1.00$_{-0.32}^{+u}$&0.82$_{-0.03}^{+1.44}$&--11.93$_{-u}^{+0.07}$&42.91$_{-0.02}^{+0.16}$\\  
      \hline
      UGC 3752&149/129&2.03$_{-0.02}^{+0.01}$&24.00$_{-0.03}^{+0.01}$&24.58$_{-0.10}^{+0.09}$&0.90$_{-0.01}^{+0.02}$&0.91$_{-0.38}^{+0.08}$&0.43$_{-0.02}^{+0.02}$&--12.22$_{-u}^{+0.05}$&42.77$_{-0.02}^{+0.01}$\\ 
      \hline
      Z319-7$^*$&133/113&1.39$_{-0.15}^{+0.17}$&23.18$_{-0.23}^{+0.14}$&-&-&-&0.05$_{-0.02}^{+0.03}$&--11.82$_{-0.09}^{+0.03}$&43.14$_{-0.09}^{+0.11}$\\  
      \hline
      Z333-49$^*$&213/192&1.67$_{-0.12}^{+0.13}$&23.20$_{-0.09}^{+0.10}$&-&-&-&0.10$_{-0.03}^{+0.04}$&--11.77$_{-0.03}^{+0.02}$&43.02$_{-0.07}^{+0.07}$\\   
      \hline
      Z367-9&196/203&1.66$_{-0.16}^{+0.13}$&23.11$_{-0.21}^{+0.12}$&23.75$_{-0.24}^{+0.13}$&0.05$_{-u}^{+u}$&0.59$_{-u}^{+u}$&0.04$_{-0.01}^{+0.02}$&--11.89$_{-0.14}^{+0.01}$&42.41$_{-0.08}^{+0.05}$\\  
      \hline
\end{tabular}
\vspace{0.3 cm}
\end{table*}
\endgroup

\begingroup
\renewcommand*{\arraystretch}{1.6}
\begin{table*}
\scriptsize
\centering
\caption{\normalsize Reanalysis best-fit results of the CT-AGN in \citet{Marchesi_2019}.\\
Col 2: $\chi^2$/d.o.f. for sources using \XMM\ and \cha\ data and cstat/d.o.f. for sources using \XRT\ data.}
\label{Table:resultsCT}
  \begin{tabular}{lccccccccccccccc}
       \hline
       \hline     
         Source&Statistics&$\Gamma$&N$\rm _{H,l.o.s}$&N$\rm _{H,tor}$&cos($\theta\rm _{inc}$)&$c_{\rm f}$&norm&Flux$_{2-10}$&L$_{\rm int,2-10}$\\
       \hline
      CGCG 164-19&38/37&1.67$_{-u}^{+0.34}$&23.77$_{-0.16}^{+0.18}$&24.56$_{-0.38}^{+u}$&0.89$_{-u}^{+u}$&0.90$_{-0.73}^{+u}$&0.05$_{-0.05}^{+0.10}$&--12.36$_{-u}^{+0.33}$&42.61$_{-0.34}^{+0.17}$\\
      \hline
      ESO 005-G004&59/47&1.90$_{-0.40}^{+u}$&24.67$_{-0.37}^{+0.37}$&24.15$_{-0.29}^{+0.28}$&0.78$_{-0.45}^{+0.15}$&0.79$_{-0.47}^{+0.14}$&0.32$_{-0.32}^{+2.78}$&--12.27$_{-u}^{+0.06}$&41.92$_{-0.23}^{+0.49}$\\
      \hline
      ESO 201-004&99/92&1.97$_{-0.35}^{+0.28}$&24.34$_{-0.14}^{+0.09}$&23.28$_{-0.12}^{+0.08}$&0.05$_{-u}^{+0.18}$&0.20$_{-u}^{+0.26}$&0.99$_{-0.63}^{+1.23}$&--12.38$_{-u}^{+0.03}$&43.90$_{-0.32}^{+0.45}$\\ 
      \hline
       MCG +08-03-018&125/131&1.98$_{-0.12}^{+0.09}$&23.97$_{-0.09}^{+0.09}$&23.00$_{-0.23}^{+0.13}$&0.55$_{-u}^{+u}$&1.00$_{-0.64}^{+u}$&0.49$_{-0.30}^{+0.22}$&--11.89$_{-0.13}^{+0.07}$&43.08$_{-0.29}^{+0.10}$\\
      \hline
       NGC 424&422/335&2.12$_{-0.04}^{+0.04}$&24.51$_{-0.02}^{+0.02}$&23.81$_{-0.07}^{+0.05}$&0.06$_{-u}^{+0.01}$&0.10$_{-u}^{+0.01}$&5.27$_{-2.76}^{+1.59}$&--12.00$_{-0.64}^{+0.12}$&43.54$_{-0.06}^{+0.06}$\\ 
      \hline
      NGC 1194&275/267&1.91$_{-0.18}^{+0.11}$&23.99$_{-0.07}^{+0.05}$&24.75$_{-0.37}^{+u}$&0.86$_{-0.16}^{+0.07}$&0.85$_{-0.20}^{+0.06}$&0.39$_{-0.20}^{+0.16}$&--11.97$_{-u}^{+0.04}$&42.71$_{-0.09}^{+0.07}$\\ 
      \hline
	NGC 3079&111/111&2.04$_{-0.22}^{+0.20}$&24.75$_{-0.37}^{+u}$&23.81$_{-0.07}^{+0.05}$&0.15$_{-0.10}^{+0.18}$&0.23$_{-0.04}^{+0.20}$&4.07$_{-4.07}^{+8.00}$&--12.28$_{-u}^{+0.44}$&42.71$_{-1.10}^{+0.45}$\\ 
      \hline
      NGC 3393&120/110&1.60$_{-0.17}^{+0.20}$&24.27$_{-0.13}^{+0.24}$&24.21$_{-0.16}^{+0.05}$&0.05$_{-u}^{+0.64}$&0.55$_{-0.42}^{+0.37}$&0.43$_{-0.20}^{+0.40}$&--12.43$_{-2.51}^{+0.18}$&42.85$_{-0.67}^{+0.80}$\\  
      \hline
      NGC 4945&1144/1167&1.92$_{-0.02}^{+0.02}$&24.55$_{-0.01}^{+0.01}$&24.65$_{-0.03}^{+0.04}$&0.05$_{-u}^{+0.01}$&0.12$_{-0.01}^{+0.01}$&76.86$_{-0.76}^{+0.76}$&--11.54$_{-0.15}^{+0.10}$&43.24$_{-0.01}^{+0.01}$\\ 
      \hline
      NGC 5100&178/171&1.69$_{-0.13}^{+0.04}$&23.29$_{-0.09}^{+0.02}$&25.50$_{-0.23}^{+u}$&0.95$_{-0.52}^{+u}$&0.25$_{-0.13}^{+0.05}$&0.17$_{-0.01}^{+0.06}$&--11.53$_{-0.09}^{+0.23}$&43.21$_{-0.04}^{+0.06}$\\  
      \hline
      NGC 5643&191/173&1.63$_{-0.14}^{+0.16}$&$>$24.24&24.07$_{-0.20}^{+0.06}$&0.95$_{-0.59}^{+u}$&0.80$_{-0.24}^{+0.11}$&0.12$_{-0.02}^{+0.03}$&--12.13$_{-2.04}^{+0.11}$&41.35$_{-0.02}^{+0.06}$\\ 
      \hline
      NGC 5728&417/446&1.80$_{-0.10}^{+0.02}$&24.09$_{-0.01}^{+0.03}$&25.21$_{-0.61}^{+u}$&0.52$_{-0.01}^{+0.21}$&0.59$_{-0.11}^{+0.00}$&1.80$_{-0.03}^{+0.92}$&--11.78$_{-0.15}^{+0.52}$&43.08$_{-0.04}^{+0.09}$\\   
      \hline
      NGC 6240&494/476&1.69$_{-0.05}^{+0.08}$&24.08$_{-0.05}^{+0.05}$&24.16$_{-0.04}^{+0.20}$&0.25$_{-u}^{+0.21}$&0.50$_{-0.21}^{+0.21}$&0.73$_{-0.15}^{+0.26}$&--11.64$_{-0.14}^{+0.23}$&43.67$_{-0.12}^{+0.15}$\\  
      \hline
      NGC 7130&81/81&1.71$_{-u}^{+0.43}$&$>$24.16&24.61$_{-0.32}^{+u}$&0.28$_{-0.17}^{+0.62}$&0.37$_{-0.23}^{+0.61}$&0.37$_{-0.33}^{+1.63}$&--12.67$_{-1.31}^{+0.06}$&42.93$_{-u}^{+0.66}$\\ 
      \hline
      NGC 7212&110/106&2.22$_{-0.29}^{+0.04}$&24.28$_{-0.11}^{+0.12}$&23.86$_{-0.19}^{+0.99}$&0.15$_{-0.06}^{+0.02}$&0.10$_{-u}^{+0.33}$&2.36$_{-0.61}^{+3.83}$&--12.25$_{-1.97}^{+0.01}$&43.78$_{-0.24}^{+0.12}$\\  
      \hline
      NGC 7479&195/165&2.29$_{-0.25}^{+0.26}$&24.76$_{-0.07}^{+0.08}$&25.08$_{-0.16}^{+u}$&0.15$_{-0.01}^{+0.01}$&0.29$_{-0.01}^{+0.01}$&46.80$_{-24.72}^{+49.59}$&--12.62$_{-2.37}^{+0.19}$&44.18$_{-0.22}^{+0.17}$\\   
      \hline
      NGC 7582&1708/1544&1.64$_{-0.03}^{+0.02}$&24.23$_{-0.04}^{+0.02}$&23.29$_{-0.06}^{+0.02}$&0.25$_{-0.06}^{+0.07}$&0.90$_{-0.11}^{+u}$&0.76$_{-0.07}^{+0.01}$ &--11.30$_{-0.06}^{+0.06}$&42.57$_{-0.01}^{+0.01}$\\  
      \hline
\end{tabular}
\vspace{0.4 cm}
\end{table*}
\endgroup

\begin{figure*} 
\begin{minipage}[b]{.5\textwidth}
\centering
\includegraphics[width=1\textwidth]{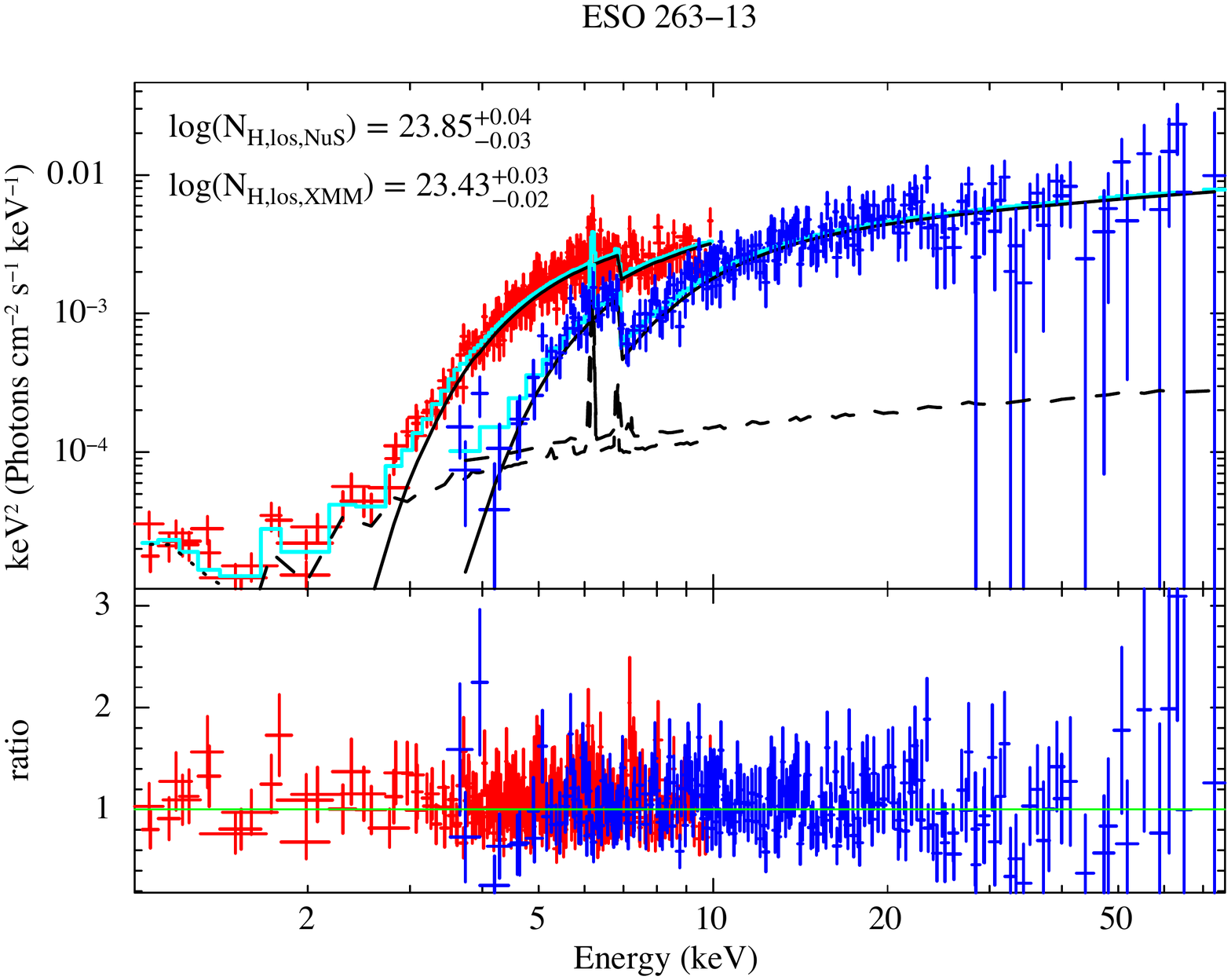}
\includegraphics[width=1\textwidth]{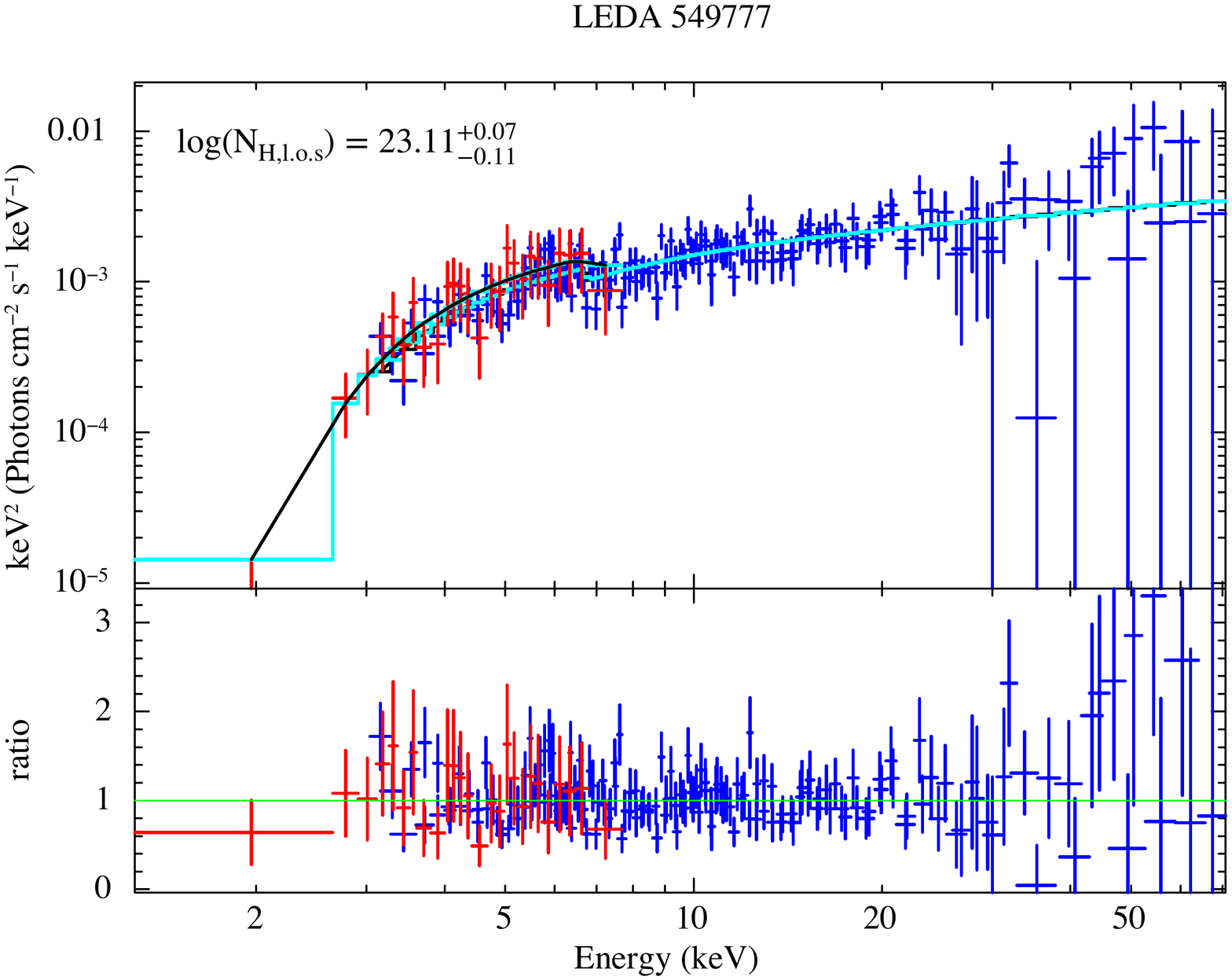}
\end{minipage}
\begin{minipage}[b]{.5\textwidth}
\centering
\includegraphics[width=1\textwidth]{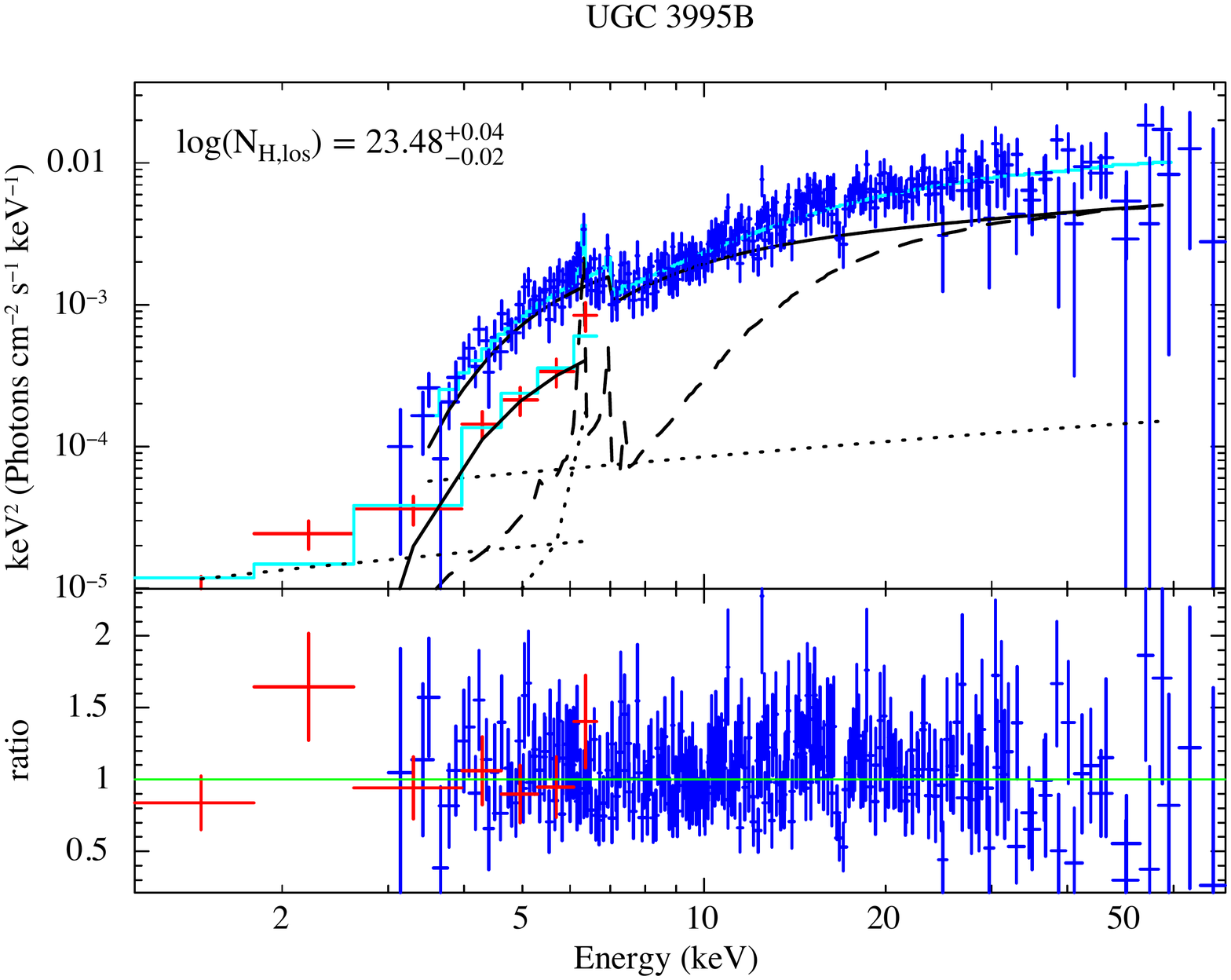}
\includegraphics[width=1\textwidth]{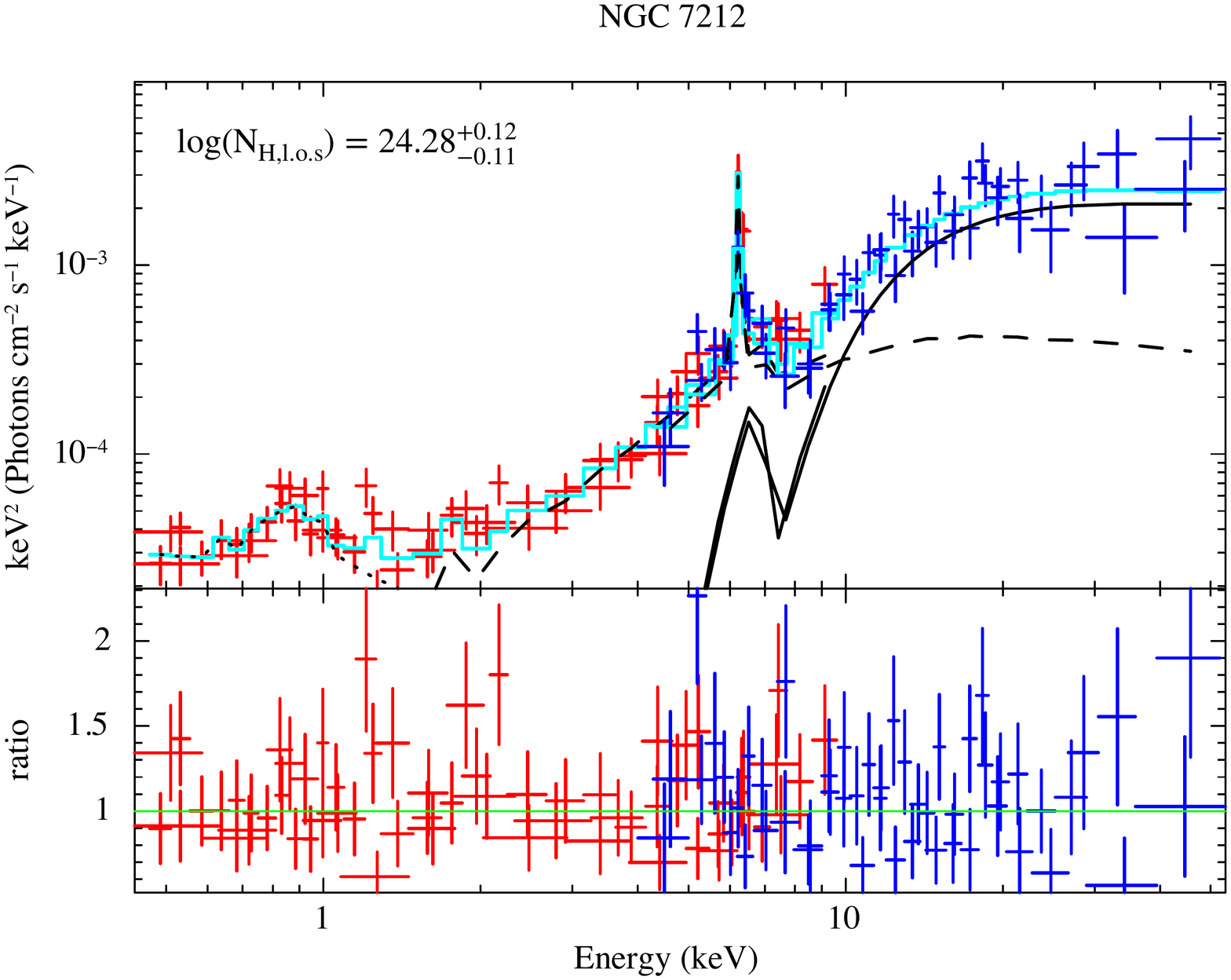}
\end{minipage}
\caption{Spectra and ratio (between the data and model predictions) of ESO 263-13 (\XMM+\NuSTAR), UCG 3995B (\cha+\NuSTAR), LEDA 549777 (\XRT+\NuSTAR) and NGC 7212 (in M19 sample). \NuSTAR\ and soft X-ray data are plotted in blue and red, respectively. The total best-fit model predictions and different components of the models of are plotted in cyan black, respectively. The reprocessed component, line-of-sight component, and scattered (\& Mekal) component are plotted in dashed, solid, and dot lines, respectively. Significant flux variabilities are found in ESO 263-13 and UGC 3995B.}
\label{fig:spectra}
\end{figure*}

\section{Variability Analysis} \label{sec:variability}
To properly characterize the spectra of the sources which are observed with flux variation between the \NuSTAR\ observations and soft X-ray observations (2--10\,keV flux variation $>$20 percent), we fit them three times: 1. allowing the cross-calibration factor between the soft X-rays spectra and \NuSTAR\ spectra, $C_{\rm soft/NuS}$, free to vary, assuming the flux variability is caused by the intrinsic emission variation. The sources whose flux variabilities are thought to be caused by intrinsic emission variations are those with $C_{\rm soft/NuS}$ is different from unity at 90\% confidence level. 2. fixing $C_{\rm soft/NuS}$ = 1, and disentangling the ``line-of-sight'' column densities of the soft X-ray observations, N$_{\rm H,los,soft}$ and \NuSTAR\ observations, N$_{\rm H,los,NuS}$, assuming the flux variability is caused by N$_{\rm H,l.o.s}$ variations. The sources whose flux variabilities are caused by N$_{\rm H,l.o.s}$ variations are those whose N$_{\rm H,los,NuS}$ are different from their N$_{\rm H,los,soft}$ at 90\% confidence level. 3. allowing $C_{\rm soft/NuS}$ free to vary and disentangling the N$_{\rm H,los,soft}$ and N$_{\rm H,los,NuS}$, assuming the flux variability is caused by both the intrinsic emission and N$_{\rm H,l.o.s}$ variations. If the improvement in the fit is $>$90\% confidence level using the \texttt{ftest} in \texttt{XSPEC} with respect to the previous two scenarios the first two scenarios, we assume that the flux variability of this source is the result of both intrinsic emission and N$_{\rm H,los}$ variations. Otherwise, the cause of the flux variation of this source is determined by comparing the reduced statistics ($\chi^2$/d.o.f or cstat/d.o.f) of the best-fit of the first two scenarios. For sources whose flux variabilities are caused by the N$_{\rm H,l.o.s}$ variations only, we fix their cross-calibration factors at $C_{\rm soft/NuS}$ = 1 to better constrain the other parameters. Here, we assume a consistent reprocessed component and scattered component between different observing epochs, considering that the global structure and properties of the obscuring torus is stable in the time-scale of few years. We summarize the best-fit results of the sources with observed flux variability (including four variable sources from M19 sample) in Table~\ref{Table:Variability}.

\begingroup
\renewcommand*{\arraystretch}{1.5}
\begin{table*}
\scriptsize
\centering
\caption{Best-fit results of sources with flux variabilities in our sample and M19 sample. Col 2: Cross-calibration factor between the soft X-ray spectra and \NuSTAR\ spectra. Col 3: Logarithm of ``line-of-sight'' column density of \NuSTAR\ observation in cm$^{-2}$. Col 4: Logarithm of ``line-of-sight'' column density of soft X-ray observation in cm$^{-2}$. Col 5: Absolute difference between N$\rm _{H,los,NuS}$ and N$\rm _{H,los,soft}$. Col 6: Time interval between the \NuSTAR\ observation and the soft X-ray observation in years. Sources labeled with asterisks are those fitted with only the line-of-sight component and the scattered component.}
\label{Table:Variability}
\vspace{.1cm}
  \begin{tabular}{l|ccccc}
       \hline
       \hline     
         Source&$C_{\rm soft/NuS}$&N$\rm _{H,los,NuS}$&N$\rm _{H,los,soft}$&$\Delta$N$\rm _{H,l.o.s}$&$\Delta$T\\
       \hline
      2MASX J06411806+3249313&0.75$_{-0.10}^{+0.12}$&23.09$_{-0.10}^{+0.08}$&23.27$_{-0.07}^{+0.06}$&0.18$_{-0.11}^{+0.12}$&12.80\\ 
      \hline
      3C 105$^*$&2.63$_{-0.19}^{+0.23}$&23.65$_{-0.03}^{+0.10}$&-&-&8.34\\ 
      \hline
      3C 445&1.43$_{-0.08}^{+0.09}$&23.08$_{-0.07}^{+0.05}$&23.21$_{-0.04}^{+0.02}$&0.13$_{-0.06}^{+0.07}$&14.44\\ 
      \hline
      3C 452&1.30$_{-0.02}^{+0.11}$&23.59$_{-0.01}^{+0.03}$&23.82$_{-0.01}^{+0.06}$&0.23$_{-0.03}^{+0.06}$&8.42\\ 
      \hline
      4C +29.30&1$^f$&23.50$_{-0.06}^{+0.04}$&23.68$_{-0.03}^{+0.03}$&0.18$_{-0.05}^{+0.07}$&5.58\\ 
      \hline
      4C +73.08&1$^f$&23.53$_{-0.12}^{+0.04}$&23.70$_{-0.05}^{+0.03}$&0.17$_{-0.06}^{+0.12}$&7.40\\ 
      \hline
      ESO 21-4$^*$&1$^f$&23.43$_{-0.10}^{+0.10}$&23.67$_{-0.13}^{+0.13}$&0.24$_{-0.16}^{+0.16}$&4.63\\ 
      \hline
      ESO 103-35&1.33$_{-0.03}^{+0.04}$&23.25$_{-0.02}^{+0.02}$&23.28$_{-0.01}^{+0.01}$&0.03$_{-0.02}^{+0.02}$&15.58\\ 
      \hline
      ESO 119-8$^*$&2.01$_{-0.30}^{+0.34}$&23.17$_{-0.09}^{+0.09}$&-&-&0.03\\ 
      \hline
      ESO 263-13&0.81$_{-0.08}^{+0.08}$&23.85$_{-0.03}^{+0.04}$&23.43$_{-0.02}^{+0.03}$&0.42$_{-0.04}^{+0.04}$&1.67\\ 
      \hline
      ESO 383-18&1.43$_{-0.02}^{+0.06}$&23.28$_{-0.02}^{+0.03}$&23.32$_{-0.03}^{+0.01}$&0.04$_{-0.04}^{+0.02}$&10.03\\ 
      \hline
      ESO 439-G009&1$^f$&23.58$_{-0.09}^{+0.09}$&23.69$_{-0.05}^{+0.07}$&0.11$_{-0.10}^{+0.11}$&0.54\\ 
      \hline
      ESO 553-43$^*$&0.63$_{-0.07}^{+0.08}$&23.18$_{-0.04}^{+0.05}$&-&-&5.75\\ 
      \hline
      Fairall 272&0.83$_{-0.09}^{+0.09}$&23.23$_{-0.05}^{+0.05}$&23.51$_{-0.06}^{+0.03}$&0.27$_{-0.08}^{+0.06}$&6.24\\ 
      \hline
      IC 1657&3.11$_{-0.83}^{+1.06}$&23.53$_{-0.21}^{+0.21}$&-&-&8.20\\ 
     \hline
      IC 4518A&0.81$_{-0.13}^{+0.15}$&23.11$_{-0.16}^{+0.12}$&23.36$_{-0.04}^{+0.07}$&0.25$_{-0.13}^{+0.17}$&6.99\\ 
      \hline
      IC 4709&2.43$_{-0.43}^{+0.49}$&23.38$_{-0.12}^{+0.02}$&-&-&6.74\\ 
      \hline
      IRAS 16288+3929&1$^f$&23.88$_{-0.10}^{+0.06}$&23.75$_{-0.11}^{+0.06}$&0.13$_{-0.12}^{+0.13}$&1.03\\ 
      \hline
      LEDA 259433&1$^f$&23.05$_{-0.24}^{+0.17}$&23.34$_{-0.28}^{+0.12}$&0.29$_{-0.29}^{+0.27}$&1.54\\
      \hline
      LEDA 2816387$^*$&1$^f$&23.76$_{-0.07}^{+0.05}$&23.87$_{-0.04}^{+0.04}$&0.11$_{-0.06}^{+0.08}$&1.54\\
      \hline
      LEDA 511869&2.17$_{-0.25}^{+0.21}$&23.88$_{-0.03}^{+0.07}$&-&-&1.71\\
      \hline
      MCG -01-05-047&1.23$_{-0.15}^{+0.16}$&23.33$_{-0.09}^{+0.09}$&23.23$_{-0.04}^{+0.03}$&0.10$_{-0.09}^{+0.10}$&3.35\\ 
      \hline    
      MCG +11-11-32&0.68$_{-0.12}^{+0.14}$&23.11$_{-0.07}^{+0.09}$&-&-&0.18\\ 
      \hline    
      Mrk 348&0.93$_{-0.03}^{+0.02}$&22.92$_{-0.03}^{+0.03}$&23.14$_{-0.01}^{+0.01}$&0.22$_{-0.03}^{+0.03}$&13.28\\ 
      \hline      
      Mrk 417&1$^f$&23.53$_{-0.07}^{+0.06}$&23.81$_{-0.05}^{+0.04}$&0.28$_{-0.08}^{+0.07}$&10.68\\ 
      \hline
      Mrk 477&1$^f$&23.30$_{-0.07}^{+0.06}$&23.70$_{-0.04}^{+0.03}$&0.40$_{-0.07}^{+0.07}$&3.82\\ 
      \hline
      Mrk 1210&1.30$_{-0.05}^{+0.05}$&23.30$_{-0.02}^{+0.04}$&-&-&11.42\\ 
      \hline
      NGC 454E&1$^f$&23.86$_{-0.04}^{+0.04}$&23.37$_{-0.03}^{+0.04}$&0.49$_{-0.06}^{+0.05}$&6.27\\ 
      \hline
      NGC 612&1.49$_{-0.08}^{+0.15}$&23.95$_{-0.04}^{+0.05}$&-&-&6.22\\ 
      \hline
      NGC 788&1.59$_{-0.10}^{+0.10}$&23.79$_{-0.05}^{+0.04}$&-&-&3.04\\ 
      \hline
      NGC 835&1$^f$&23.46$_{-0.06}^{+0.09}$&24.06$_{-0.08}^{+0.14}$&0.60$_{-0.12}^{+0.17}$&15.64\\ 
      \hline
      NGC 1142&1.43$_{-0.37}^{+0.39}$&24.20$_{-0.07}^{+0.08}$&23.75$_{-0.05}^{+0.07}$&0.45$_{-0.10}^{+0.09}$&11.71\\ 
      \hline
      NGC 3281&0.65$_{-0.07}^{+0.07}$&24.30$_{-0.06}^{+0.06}$&23.90$_{-0.04}^{+0.05}$&0.40$_{-0.08}^{+0.07}$&5.05\\ 
      \hline
      NGC 4388&1.89$_{-0.04}^{+0.05}$&23.43$_{-0.02}^{+0.02}$&-&-&5.70\\ 
      \hline
      NGC 4507&0.68$_{-0.00}^{+0.01}$&23.88$_{-0.01}^{+0.01}$&-&-&4.85\\ 
      \hline
      NGC 4939&1$^f$&23.82$_{-0.08}^{+0.09}$&23.69$_{-0.08}^{+0.09}$&0.13$_{-0.12}^{+0.12}$&6.69\\ 
      \hline
      NGC 4941&1$^f$&23.98$_{-0.20}^{+0.30}$&23.58$_{-0.24}^{+0.32}$&0.40$_{-0.38}^{+0.38}$&6.69\\ 
      \hline
      NGC 4992&1.12$_{-0.10}^{+0.11}$&23.46$_{-0.03}^{+0.03}$&23.67$_{-0.01}^{+0.03}$&0.21$_{-0.03}^{+0.04}$&10.58\\ 
      \hline
      NGC 5283&0.64$_{-0.07}^{+0.07}$&23.04$_{-0.09}^{+0.11}$&-&-&14.98\\ 
      \hline
      NGC 6300&0.085$_{-0.006}^{+0.008}$&23.21$_{-0.05}^{+0.03}$&23.34$_{-0.01}^{+0.03}$&0.13$_{-0.03}^{+0.06}$&15.48\\ 
      \hline
      NGC 7319&3.33$_{-0.20}^{+0.24}$&23.81$_{-0.05}^{+0.04}$&-&-&15.80\\ 
      \hline
      UGC 3995B&0.30$_{-0.04}^{+0.04}$&23.48$_{-0.02}^{+0.04}$&-&-&0.75\\ 
      \hline 
      Z367-9&4.03$_{-0.46}^{+0.50}$&23.11$_{-0.21}^{+0.12}$&-&-&7.56\\ 
      \hline
      CGCG 164-19&1$^f$&23.77$_{-0.16}^{+0.18}$&23.41$_{-0.35}^{+0.27}$&0.36$_{-0.31}^{+0.39}$&2.80\\ 
      \hline
      ESO 201-004&1$^f$&24.34$_{-0.14}^{+0.09}$&24.19$_{-0.14}^{+0.08}$&0.15$_{-0.15}^{+0.17}$&7.07\\ 
      \hline
      NGC 7479&0.78$_{-0.11}^{+0.14}$&24.76$_{-0.07}^{+0.08}$&24.56$_{-0.06}^{+0.04}$&0.20$_{-0.08}^{+0.10}$&2.05\\ 
      \hline
      NGC 7582&0.56$_{-0.01}^{+0.06}$&24.23$_{-0.04}^{+0.02}$&23.58$_{-0.01}^{+0.02}$&0.65$_{-0.04}^{+0.02}$&3.66\\ 
      \hline
      \hline
\end{tabular}
\vspace{0.2cm}
\end{table*}
\endgroup

\end{appendix}
\end{document}